\documentclass[12pt,draftclsnofoot,comsoc,onecolumn]{IEEEtran}

\usepackage[english]{babel}
\usepackage[utf8x]{inputenc}
\usepackage[T1]{fontenc}
\usepackage{float}
\usepackage{graphicx}
\usepackage{caption}
\usepackage[ruled,linesnumbered]{algorithm2e}
\usepackage{amsmath,amssymb,amsthm}
\usepackage{booktabs}
\usepackage{setspace}
\usepackage{xcolor}
\usepackage{comment}
\usepackage{subcaption}
\usepackage{multirow}
\usepackage{amsfonts}
\usepackage{url}

\newtheorem{theorem}{Theorem}

\newcommand\figSizeOne{0.49}
\newcommand\figSizeTwo{0.24}

\allowdisplaybreaks

\begin{document}

	\title{Energy-Efficient Proactive Caching with Multipath Routing}
		
	\author{
		Yantong Wang, Vasilis Friderikos
		\thanks{
			\IEEEcompsocthanksitem Yantong Wang and Vasilis Friderikos are with the Center of Telecommunication Research, Department of Engineering, King's College London, London, WC2R 2LS, UK. E-mail: \protect\{yantong.wang, vasilis.friderikos\}@kcl.ac.uk.
		}
	}

	\maketitle
		
	\begin{abstract}
	    The ever-continuing explosive growth of on-demand content distribution has imposed great pressure on mobile/wireless network infrastructures. To ease congestion in the network and to increase perceived user experience, caching of popular content closer to the end-users can play a significant role and as such this issue has received significant attention over the last few years. Additionally, energy efficiency is treated as a fundamental requirement in the design of next-generation mobile networks. However, there has been little attention to the overlapping area between energy efficiency and network caching especially when considering multipath routing. To this end, this paper proposes an energy-efficient  caching with multipath routing support. The proposed scheme provides a joint anchoring of popular content into a set of potential caching nodes with optimized multipath support whilst ensuring a balance between transmission and caching energy cost. The proposed model also considers different content delivery modes, such as multicast and unicast. Two separated Integer-Linear Programming (ILP) models are formulated for each delivery mode. To tackle the curse of dimensionality we then provide a greedy simulated annealing algorithm, which not only reduces the time complexity but also provides a competitive performance. 
       	A wide set of numerical investigations reveal that the proposed scheme reduces the energy consumption up to $80\%$ compared with other widely used caching approaches under the premise of network resource limitation. Sensitivity analysis to different parameters is also meticulously discussed in this paper. 
	\end{abstract}
	
	\begin{IEEEkeywords}
		Energy Efficiency, Integer Linear Programming, Multipath Routing, Proactive Caching, Simulated Annealing, Wireless Networks
	\end{IEEEkeywords}

\section{Introduction}
\label{sec:intro}

The enormous amount of mobile traffic, which is expected to reach more than 300 exabytes per month in 2026~\cite{cerwall2021ericsson}, is creating great challenges in the current and emerging mobile network infrastructures. 
As an evolutionary network architecture, Information-Centric Networking (ICN) is designed to support content distribution and user mobility with in-network caching feature~\cite{xylomenos2014survey}. Compared with other caching approaches, proactive caching methods make a trade-off between storage capacity and latency, whose key aspects are related to the following type of questions: "where to cache", "what to cache", "cache dimensioning", and "content delivery"~\cite{wang2020survey}. In this paper, the focus is on the location to anchor popular content and determining an efficient routing path to be employed for caching delivery. Despite the well-known benefits of mobile/wireless ICN, there are still a number of significant and challenging open-ended topics in this field, such as utilizing multipath routing, security and energy efficiency~\cite{fang2018survey}. When addressing energy consumption, it is treated as one of the key pillars in the design of 5G and beyond networks, which is not only important in monetary terms due to energy savings but also for curbing the $CO_2$ emissions~\cite{buzzi2016survey} and aligning operators with the goals related to the so-called 1.5$^{\circ}$C climate trajectory\footnote{\url{www.ipcc.ch/sr15/chapter/spm}}.
Recently, 5G has been commercialized in many markets across the world and 6G has already started to attract significant attentions from both academia and industry~\cite{dogra2020survey}. Though the vision and requirements of 6G are still under study, there is no doubt that reducing the overall network energy consumption is a significant target for 6G. According to~\cite{latva2019key}, the energy efficiency of 6G is expected to be improved by at least a factor of 10, when compared with 5G networks. 
The energy consumption in ICN can be divided into two parts~\cite{choi2012network,llorca2013dynamic,fang2015energy,dehghani2018ccn,li2013energy,chen2018joint,li2020energy}: the energy required for content caching and the energy for data transmission. 
Content replicas can be placed near access routers as a mean of reducing transmission costs but with an extra cost paid for hosting popular content; reversely, consolidating caching host nodes can be adopted to reduce caching energy consumption at the expense of an increased transmission cost. This inherent trade-off raises the following key question: where to cache popular content and how to design an efficient routing path to minimize the total energy consumption whilst satisfying the network resource limitations.

There are a number of prior works on the issues related to location selection for hosting popular content in order to minimize energy consumption. An integer linear programming (ILP) model is studied in~\cite{choi2012network}, where the limitation of router caching space and different caching hardware technologies have been taken into account. J. Llorca \textit{et al.}~\cite{llorca2013dynamic} explicitly consider the temporal and spatial dynamics of user requests and the heterogeneity of network resources. In~\cite{fang2015energy}, the authors present a distributed in-network caching solution, in which each content router only needs local information. An energy consumption model with propagation delay-awareness is investigated in~\cite{dehghani2018ccn} and a quantized Hopfield neural network is proposed to derive competitive solutions. The relationship between aggregate energy consumption per transmitted bit of information and the average number of hops has been studied in~\cite{li2013energy}. A hybrid caching policy for reducing latency and energy consumption is designed in~\cite{xu2019energy} which consists of device local caching, device-to-device (D2D) caching, base station (BS) caching. 
Except for saving the traditional energy in the above works, energy harvesting technology together with cooperative caching has also received scholarly attention in recent years, such as BS caching in~\cite{zhang2020sleeping,vallero2021base} and D2D caching in~\cite{meng2021cache}. In particular, the authors in~\cite{zhang2020sleeping} formulate a BS caching problem as a mixed integer linear programming model and then decompose it as two sub-problems, and the work~\cite{meng2021cache} investigates the caching performance with stochastic geometry. Unlike~\cite{zhang2020sleeping} and~\cite{meng2021cache}, the study in~\cite{vallero2021base} uses a simulation-based method, instead of an optimization model, to measure the energy efficiency of different BSs switching. Furthermore, a number of studies have investigated the energy-efficient schemes by jointly considering caching, communication, and computation. For instance, the authors in~\cite{chen2018joint} propose a framework combining software defined networking, caching, and computing to balance the energy consumption and network resource utilization, which has the potential to support general in-network services. The inherent lack of user request arrival knowledge case is considered in \cite{li2020energy} where caching, transcoding, and backhaul retrieving are jointly optimized. Another joint optimization research can be found in~\cite{zafari2018optimal} which concentrates on the trade-offs between communication, computation and caching. The authors also provide a $\epsilon$-optimal solution to the proposed non-linear optimization model. 

However, the majority of the aforementioned works do not take the routing policies into consideration~\cite{choi2012network,fang2015energy,dehghani2018ccn,li2020energy,xu2019energy,zhang2020sleeping,vallero2021base,meng2021cache} or only adopt the shortest path~\cite{llorca2013dynamic,li2013energy,chen2018joint,zafari2018optimal}. It is insufficient to only consider the shortest path routing especially when the network bandwidth is limited. There are also some researches discussing the routing design in caching policies. For example, the study in \cite{liu2019joint} formulates the joint caching and multipath routing problem as an ILP model to maximize the volume of caching-served file requests in the wireless network. Chu \textit{et al.}~\cite{chu2018joint} jointly consider the cache allocation and routing path to maximize content utilities based on cache partitioning. The delay optimization scenario is also considered in~\cite{chu2018joint}. The authors in~\cite{ioannidis2018jointly} investigate the caching and routing problem considering both source routing and nominal shortest path routing
to minimize routing cost. To simplify the model, the link bandwidth in \cite{ioannidis2018jointly} is assumed to be abundant. However, none of the above studies~\cite{liu2019joint,chu2018joint,ioannidis2018jointly} consider caching placement for energy efficiency purposes. For a more detailed and extensive treatment of the overall state-of-the-art caching and routing policies in ICN, we refer the reader to~\cite{seetharam2017caching}.

In this article, we jointly consider the caching placement and multipath routing to minimize the total energy consumption and study the following two content delivery modes~\cite{liu2016caching}: 1) the service flow can be shared if the users have the same content preference as well as connected access router, which is called \textit{multicast mode}. For example, semi-synchronous requests can be bundled together and served via a single transmission; 2) the service flow cannot be shared and each request is satisfied with an independent (unicast) transmission, i.e. \textit{unicast mode}. We also consider explicitly the caching memory space and link bandwidth limitations. 
In the structure of integrated multi-access edge computing (MEC) platform and 5G network~\cite{etsi2020mec}, the proposed caching scheme can be deployed in the MEC orchestrator. Moreover, the transmission mode, either unicast or multicast, is managed by the session management function and user plane function, which control user-plane packet forwarding in 5G core network~\cite{etsi2018fifthg}.
The existing research in~\cite{ayoub2018energy} is closest to the problem we tackle here. Though~\cite{ayoub2018energy} considers the energy-efficient caching and routing algorithm, here are some main differences between our work and~\cite{ayoub2018energy}. On the one hand, the caching methods are quite different. For example, we allow replicated caching while~\cite{ayoub2018energy} considers unique caching that each content is satisfied by exactly one node. 
On the other hand, our work includes modelling for different delivery modes, i.e. multicast and unicast, while~\cite{ayoub2018energy} only considers the nominal method of delivery which is the unicast.
Based on the aforementioned closely relevant state of the art, the main contributions of this work can be summarized as follows.
\begin{itemize}
	\item An energy-efficient caching scheme is proposed by jointly considering the content placement and multipath routing.  The problem is  formulated as an optimization model with caching space and link capacity limitations. In particular, two ILP models are detailed according to the two delivery modes, multicast and unicast respectively.
	\item Since the proposed ILP models belong to the $NP$-hard family, we provide a heuristic algorithm named Greedy Simulated Annealing Caching (GSAC). In this approach, we divide the original problem into several sub-problems and employ the simulated annealing to find sub-optimal, albeit competitive, solutions for caching nodes recursively.
	\item To understand the performance gap among evaluated methods, a wide-set of simulations is performed under different scenarios. More specifically, we evaluate influences from the aspects of the user (in terms of the number of requests, user preference and prediction accuracy) and the network (such as the available bandwidth and caching space and the underlying network topology). 
\end{itemize}

The rest of this paper is organized as follows. Section \ref{sec:model} formulates the system model. Section \ref{sec:heuristic} provides a simulated annealing-based heuristic algorithm. In the end, Section \ref{sec:numerical} and \ref{sec:conclusions} give the numerical investigations and conclusions respectively. 
\section{System Model}
\label{sec:model}

\subsection{Network Model}
\label{subsec:network_model}
\begin{figure}[!htbp]
	\centering
	\includegraphics[width=.82\textwidth]{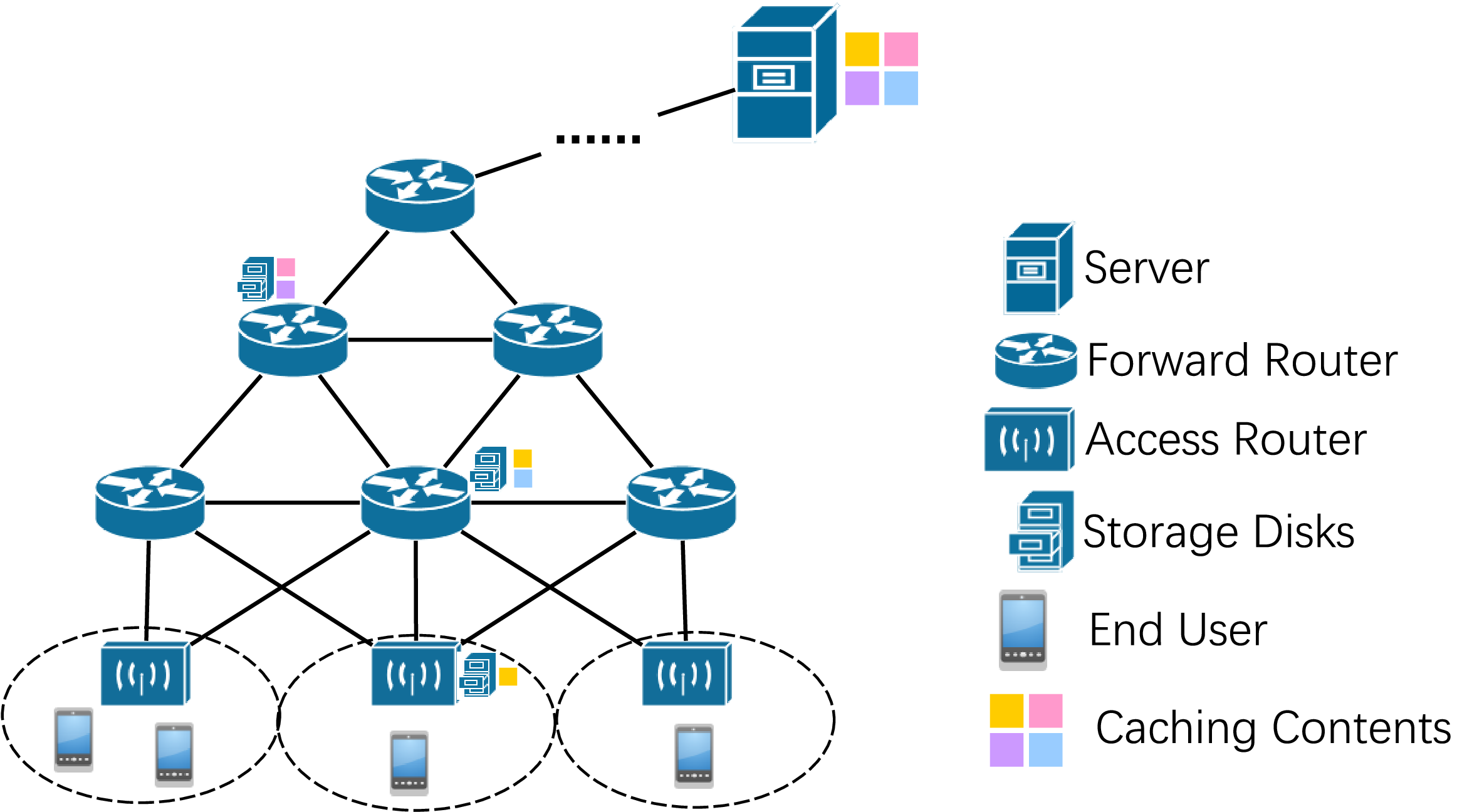}
	\caption{Network topology, network entities and overall system model}
	\label{fig:system}
\end{figure}

The network is modelled as an undirected graph $\mathcal{G}=\{\mathcal{E},\mathcal{L}\}$ as shown in Figure \ref{fig:system}, where $\mathcal{E}$ represents the set of nodes and $\mathcal{L}$ denotes the set of links. Each link $l\in\mathcal{L}$ is equipped with bandwidth $c_l$. The nodes are partitioned to access routers (AR) $\mathcal{A}\subset\mathcal{E}$, forward routers $\mathcal{F}\subset\mathcal{E}$ and servers $\{0\}\in\mathcal{E}$. Additionally, each caching node $e\in\mathcal{E}^0=\mathcal{E}\backslash\{0\}$ has storage capacity $w_e$ and can provide both data flow forwarding and content caching services.

Let $\mathcal{N}$ be the set of contents. For each content $n\in\mathcal{N}$, the required caching size and transmission bandwidth are represented by $s_n$ and $b_n$ respectively. Without loss of generality, we assume all contents can be fetched from the backbone servers in case of a cache miss. The user's preference is described as $\rho_{kn}$, which is the probability of user $k$ requesting file $n$. Furthermore, we use $\pi_{ka}$ to represent the user $k$ accesses to the network service via AR $a$, which depends on the user movement behaviour. For modelling simplification purposes, we assume the user's movement is already known, thus $\pi_{ka}$ becomes a binary indicator; otherwise, $\pi_{ka}$ is the moving probability and the model for this case is discussed in our previous work~\cite{wang2019proactive}. Consequently the aggregated number of request for content $n$ at AR $a$, $\lambda_{na}$, can be determined by $\lambda_{na}=\sum_{k}\pi_{ka}\cdot\rho_{kn}$.
For ease of reference, all used symbols and their meanings are summarized in Table \ref{tab:notations}.

\begin{table}[!htbp]
\centering
\caption{Summary of Main Notations}
\label{tab:notations}
\small{\singlespacing{
	\begin{tabular}{cl}
	\toprule
	\textbf{Symbol}&\textbf{Description}\\
	\midrule
	$\mathcal{E}^0$ & the set of caching candidates \\
	$\mathcal{A}$ & the set of access routers (AR), $A\subset\mathcal{E}^0$\\
	$\{0\}$ & the data server \\
	$\mathcal{N}$ & the set of contents \\
	$\mathcal{E}$ & the set of nodes, $\mathcal{E}=\mathcal{E}^0\cup\{0\}$\\
	$\mathcal{L}$ & the set of links \\
	$\mathcal{K}$ & the set of users \\
	\midrule
	$\alpha$ & power efficiency of caching 1 bit data (Watts per bit)\\
	$\beta$ & power cost for transmitting 1 bit data within 1 hop \\
	& (Joule per bit per hop) \\
	$T$ & the time interval normalization factor \\
	$s_n$ & the size of content $n$ \\
	$b_n$ & the bandwidth requirement of content $n$\\
	$w_e$ & the available space in caching node $e$ \\
	$c_l$ & the capacity of link $l$ \\
	$\rho_{kn}$ & the preference on content $n$ for user $k$\\
	$\pi_{ka}$ & the connection with AR $a$ for user $k$\\
	$\lambda_{na}$ & the aggregated number of request for content $n$ at AR $a$\\
	$N_{aep}$ & the number of hops for path $p\in\mathcal{P}_{ae}$\\
	$B_{laep}$ & indicates the link $l$ is in path $p\in\mathcal{P}_{ae}$ or not\\
	\midrule
	$x_{ne}$ & indicates the content $n$ is cached in node $e$ or not\\
	$y_{naep}$ & indicates the delivery of content $n$ uses path $p\in\mathcal{P}_{ae}$ or not\\
	\bottomrule
	\end{tabular}	
}}
\end{table}

\subsection{Energy Model}
\label{subsec:energy_model}
In this paper, we focus on two types of energy consumption in the caching-enabled network: the caching energy $E^C$ and the communication transmission energy $E^T$. Therefore, the total energy consumption can be represented by $E=E^C+E^T$. Similar to \cite{li2020energy}, the caching energy is a proportional function of content size and can be given by, 
\begin{equation}
	E^C=\sum_{n\in\mathcal{N}}\sum_{e\in\mathcal{E}^0}\!\alpha\!\cdot\!s_n\!\cdot\!x_{ne}\!\cdot\!T,
\end{equation} 
where $\alpha$ is the power efficiency for caching 1-bit of information; $s_n$ is the required caching space for content $n$; $x_{ne}$ is a binary decision variable with,

\hangafter 1
\hangindent 2em
\noindent
\\$ x_{ne}\!=\!
\begin{cases}
	1,  &\text{if content $n$ is cached in node $e$;}  \\
	0,  & \text{otherwise.}\\
\end{cases}$
\vspace{1em}

\noindent{$T$ is the time interval normalization factor partitioned in time epochs of equal length and the hosting content in one time period might be totally different from the data in the next time epoch; this is also called as caching refreshing cycle~\cite{li2020energy} or windowed grouped data aggregation~\cite{zafari2018optimal}. Without loss of generality, hereafter we focus on optimization procedures in one time epoch.}

The communication transmission energy consumption is measured by the number of hops in the network and can be given by, 
\begin{equation}
	E^T=\sum_{n\in\mathcal{N}}\sum_{a\in\mathcal{A}}\sum_{e\in\mathcal{E}}\sum_{p\in\mathcal{P}_{ae}}\!\beta\!\cdot\!N_{aep}\!\cdot\!s_n\!\cdot\!y_{naep},
\end{equation}
where $\beta$ is the energy cost for transmitting 1-bit of information within a single hop; $N_{aep}$ expresses the number of hops on path $p\in\mathcal{P}_{ae}$ and $\mathcal{P}_{ae}$ represents all available paths between AR $a$ and caching node $e$; notably $N_{aep}=0$ if $a=e$; $y_{naep}$ is a decision variable with 

\hangafter 1
\hangindent 2em
\noindent
\\$ y_{naep}\!=\!
\begin{cases}
	1,  &\text{if the delivery of content $n$ uses path $p\in\mathcal{P}_{ae}$;}  \\
	0,  & \text{otherwise.}\\
\end{cases}$
\vspace{1em}

It is worth noting that we do not take the static energy consumption of network devices into consideration, such as the idle power to support routers and network interfaces. The rational behind this is that in this paper we assume that these devices are being kept in power-on mode to provide various kinds of network services for different requests, instead of only content caching.

\subsection{Optimization Model}
\label{subsec:optimization_model}

In this subsection, we consider two types of content delivery mode, i.e. the multicast and the unicast mode. For modelling simplification purposes, the following optimization models are under the assumption that the user's interested content is already known, which indicates $\rho_{kn}$ becomes a binary indicator to bridge user $k$ and content $n$. The user preference prediction has been extensively studied in the context of caching, such as \cite{chen2018caching} and \cite{ren2019incentivized}, and this technique is beyond the scope as well as complimentary of this paper. 
The influence of the predicted accuracy on system energy consumption is discussed in Section \ref{sec:numerical}. 
Moreover, in this paper we focus on delay-insensitive requests, such as non real-time HTTP video downloading and video on demand services. Therefore, the latency constraint for energy-efficient caching is not considered in the following models and would be studied as a future work.

\subsubsection{Multicast Mode}

The ILP model is formulated as follows: 
\begin{subequations}
\label{fml:OP1}
\begin{align}
	\label{fml:OP1-obj}
	(P1)\qquad \mathop{\min}_{x_{ne},y_{naep}}\;&E\\
	\textrm{s.t.}\qquad
	\label{fml:OP1-con1}
	&\sum_{n\in\mathcal{N}}\!s_n\!\cdot\!x_{ne}\!\leq\!w_e,\forall e\in\mathcal{E}^0;\\
	\label{fml:OP1-con2}
	&\sum_{n\in\mathcal{N}}\sum_{a\in\mathcal{A}}\sum_{e\in\mathcal{E}}\sum_{p\in\mathcal{P}_{ae}}\!b_n\!\cdot\!B_{laep}\!\cdot\!y_{naep}\!\leq\!c_l,\forall l\in\mathcal{L};\\
	\label{fml:OP1-con3}
	&y_{naep}\!\leq\!x_{ne},\forall n\in\mathcal{N},a\in\mathcal{A},e\in\mathcal{E}^0,p\in\mathcal{P}_{ae};\\
	\label{fml:OP1-con4}
	&y_{naep}\!\leq\!\lambda_{na},\forall n\in\mathcal{N},a\in\mathcal{A},e\in\mathcal{E},p\in\mathcal{P}_{ae};\\
	\label{fml:OP1-con5}
	&\lambda_{na}\!\cdot\!\sum_{e\in\mathcal{E}}\sum_{p\in\mathcal{P}_{ae}}\!y_{naep}\!=\!\lambda_{na},\forall n\in\mathcal{N},a\in\mathcal{A};\\
	\label{fml:OP1-con6}
	&x_{ne}\in\{0,1\},\forall n\in\mathcal{N},e\in\mathcal{E}^0;\\
	\label{fml:OP1-con7}
	&y_{naep}\in\{0,1\},\forall n\in\mathcal{N},a\in\mathcal{A},e\in\mathcal{E},p\in\mathcal{P}_{ae}.
\end{align}
\end{subequations}
The objective function \eqref{fml:OP1-obj} aims to minimize the total energy consumption, which is introduced in subsection \ref{subsec:energy_model}. Constraint \eqref{fml:OP1-con1} relates to the storage space used for caching which should not exceed the node capacity. Constraint \eqref{fml:OP1-con2} satisfies the link bandwidth limitation. We note that $B_{laep}$ is a binary indicator, where $B_{laep}=1$ means that the link $l$ is in path $p\in\mathcal{P}_{ae}$, and $B_{laep}=0$, otherwise. Therefore, $\sum_a\sum_e\sum_p B_{laep}\cdot y_{naep}$ selects all links used for content $n$ delivery. Constraint \eqref{fml:OP1-con3} limits the source of delivery path should host related contents. Moreover, in constraint \eqref{fml:OP1-con4} when $\lambda_{na}=0$, $y_{naep}$ is forced to be $0$, which means the delivery path to AR $a$ should not be assigned if there is no such content requirement in that AR; and in constraint \eqref{fml:OP1-con5} if $\lambda_{na}\neq 0$, $\sum_e\sum_p y_{naep}=1$ indicating that when there is content requirement in AR $a$ we must allocate a delivery path accordingly. It is worth noticing that when dealing with caching allocation, we only focus on the storage node $e\in\mathcal{E}^0$ such as constraints \eqref{fml:OP1-con1} and \eqref{fml:OP1-con3}; however, for the path assignment, we consider both the caching node $e\in\mathcal{E}^0$ (for cache hitting case) and the core network server $\{0\}$ (for the cache missing case), like constraint \eqref{fml:OP1-con2}, \eqref{fml:OP1-con4} and \eqref{fml:OP1-con5}.

\subsubsection{Unicast Mode} 

In the unicast mode of delivery, $y_{naep}$ is no longer binary but an integer decision variable where the value represents the number of flows for content $n$ that traverse path $p\in\mathcal{P}_{ae}$. In that case,  the proposed ILP model would be as follows,
\begin{subequations}
\label{fml:OP2}
\begin{align}
	\label{fml:OP2-obj}
	(P2)\qquad \mathop{\min}_{x_{ne},y_{naep}}\;&E\\
	\textrm{s.t.}\qquad
	\label{fml:OP2-con1}
	&y_{naep}\!\leq\!M\!\cdot\!x_{ne},\forall n\in\mathcal{N},a\in\mathcal{A},e\in\mathcal{E}^0,p\in\mathcal{P}_{ae};\\
	\label{fml:OP2-con2}
	&\sum_{e\in\mathcal{E}}\sum_{p\in\mathcal{P}_{ae}}\!y_{naep}\!=\!\lambda_{na},\forall n\in\mathcal{N},a\in\mathcal{A};\\
	\label{fml:OP2-con3}
	&y_{naep}\in\mathbb{Z}^{0+},\forall n\in\mathcal{N},a\in\mathcal{A},e\in\mathcal{E},p\in\mathcal{P}_{ae};\\
	&\eqref{fml:OP1-con1},\eqref{fml:OP1-con2},\eqref{fml:OP1-con6}. \nonumber
\end{align}
\end{subequations}
The constraint \eqref{fml:OP2-con1} updates from constraint \eqref{fml:OP1-con3} and $M$ is a sufficiently large number, which represents this node can serve the requests as long as it hosts the required content. For the scenario considering node service capability, $M$ should be replaced by the service rate, i.e. the number of requests that a node can serve during the time period $T$, which is discussed in~\cite{wang2019proactive}. Constraint \eqref{fml:OP2-con2} replaces constraint \eqref{fml:OP1-con4} and \eqref{fml:OP1-con5}, where all served flows for content $n$ at AR $a$ should exactly match the number of requests.

Regarding the time complexity of aforementioned models, we have the following theorem:
\begin{theorem}
    \label{thm:np}
    The ILP models $(P1)$ and $(P2)$ fall into the family of $NP$-hard problems.
\end{theorem}
\begin{proof}
    For $(P1)$, we will establish the proof by showing that model \eqref{fml:OP1} is a generalization of the uncapacitated facility location (UFL) problem, where UFL is a well-known $NP$-hard problem. Suppose the network is a tree topology and all users request the same content, then the dimension $p$ and $n$ in decision variables can be reduced, which become $x_e$ and $y_{ae}$. Consider the case that the caching space and link bandwidth are infinity, i.e. $w_e=c_l=+\infty$, and, hence, constraints \eqref{fml:OP1-con1} and \eqref{fml:OP1-con2} can be relaxed. Furthermore, let $\lambda_a\geq 1,\forall a\in\mathcal{A}$. As a result, constraint \eqref{fml:OP1-con4} is guaranteed by \eqref{fml:OP1-con7}, and $\lambda_a$ can be safely divided on both sides of \eqref{fml:OP1-con5}. In the end, (P1) can be rewritten as a UFL problem as follows: 
    \begin{subequations}
        \begin{align}
	        \mathop{\min}_{x_{ne},y_{naep}}\;&\sum_{e\in\mathcal{E}^0}(\alpha\!\cdot\!s\!\cdot\!T)x_{e}\!+\!\sum_{a\in\mathcal{A}}\sum_{e\in\mathcal{E}}(\beta\!\cdot\!s\!\cdot\!N_{ae})y_{ae}\\
	        \textrm{s.t.}\qquad
        	&y_{ae}\!\leq\!x_{e},\forall a\in\mathcal{A},e\in\mathcal{E}^0;\\
        	&\sum_{e\in\mathcal{E}}y_{ae}\!=\!1,\forall a\in\mathcal{A};\\
        	&x_{e}\in\{0,1\},\forall e\in\mathcal{E}^0;\\
            &y_{ae}\in\{0,1\},\forall a\in\mathcal{A},e\in\mathcal{E}.
        \end{align}
    \end{subequations}
    This means $(P1)$ is at least as difficult as UFL. Thus, we have proved that $(P1)$ is $NP$-hard.
    
    Similarly, by setting $\lambda_{na}=1$ in constraint \eqref{fml:OP2-con2} and slacking model \eqref{fml:OP2} as above, $(P2)$ can be converted to UFL optimization model, which means $(P2)$ is a generalization of the UFL problem, and consequently $(P2)$ is $NP$-hard as well. 
\end{proof}

Therefore, due to the curse of dimensionality it is challenging to solve problems $(P1)$ and $(P2)$ in a reasonable time when the network instances becomes large. In order to accelerate the computational time and provide decision making that is amenable for real-time operation we propose a Greedy Simulated Annealing Caching (GSAC) algorithm in the next section.

\section{A Metaheuristic Optimization Algorithm}
\label{sec:heuristic}

In GSAC, the original optimization model is decomposed by the content categories and these contents are allocated in descending order, i.e. the network serves the content with the most requests first. Then, for each content assignment, we employ a simulated annealing algorithm to allocate popular content in the most suitable nodes. 

Simulated Annealing (SA) \cite{Pardalos2002} is a well-known optimization method for approximating the global optimal solution.  Generally speaking, SA is an approach inspired by statistical mechanics concepts, and it is motivated by an analogy with the behaviour of physical systems during the cooling process. It can be deemed as a powerful heuristic method which is suitable for global optimization which requires no particular assumption on the structure of the objective function.
The main structure of SA contains two nested loops, where the external is controlled by the temperature and the internal is managed by a Markov-chain length. In each loop a new solution is produced which is then compared with the previously stored result. The new solution is accepted if its performance is better than the previous; otherwise, the new solution would be accepted with some predefined probability. The iterations continue until meeting the stopping criterion. 

Regarding the optimization model $(P1)$, the solution generated in each loop includes the decision variable $x_{e}$ and $y_{aep}$ (here we deal with the sub-problem oriented towards specific content $n$, so $n$ is omitted from the  subscript notation for simplification). Note, however, that the decision about those two variables, i.e.,  $x_{e}$ and $y_{aep}$, might result in an infeasible allocation. To restore feasibility, we set $x_{e}$ as the pivot variable. Then, once $x_{e}$ is allocated, $y_{aep}$ can be determined as follows,
\begin{equation}
	\label{fml:det_y1}
	\begin{cases}
		y_{a\{0\}1}=1,  &\text{if $\sum_{e\in\mathcal{E}^0}x_{e}=0$ and $\{a|\lambda_{a}>0\}$;}  \\
		y_{aep}=1,  &\text{if $\{(a,e,p)|\lambda_{a}>0,x_{e}=1,\min N_{ae1}, b\leq c_l, \forall a\in\mathcal{A},$}\\
		&\quad\text{$e\in\mathcal{E}^0,l\in p\in\mathcal{P}_{ae}\}$;}\\
		y_{aep}=0, &\text{otherwise.}
	\end{cases}
\end{equation}
When there is a cache miss ($\sum_{e\in\mathcal{E}^0}x_{e}=0$), the request is redirected to the backbone server $\{0\}$ via the shortest path (i.e. $p=1$); when some nodes host the content, the delivery path is from the requesting AR ($\lambda_{a}>0$) to the nearest caching node ($\min N_{ae1},x_e=1$). Additionally, the link bandwidth constraint is also considered in routing by setting $b\leq c_l$; otherwise, $y_{aep}=0$. 

The determination of $y_{aep}$ in problem $(P2)$ is as follows:
\begin{equation}
	\label{fml:det_y2}
	\begin{cases}
		y_{a\{0\}1}=\lambda_{a},  &\text{if $\sum_{e\in\mathcal{E}^0}x_{e}=0$ and $\{a|\lambda_{a}>0\}$;}  \\
		y_{aep}=\lfloor c_l/b\rfloor,  &\text{if $\{(a,e,p)|\lambda_{a}>0,x_{e}=1,\min N_{ae1}, b\leq c_l, \forall a\in\mathcal{A},$}\\
		&\quad\text{$e\in\mathcal{E}^0,l\in p\in\mathcal{P}_{ae}\}$;}\\
		y_{aep}=0, &\text{otherwise.}
	\end{cases}
\end{equation}
Similar to the multicast case, all requests are served at the backbone server $\{0\}$ in the cache-miss case; when some nodes host the content, the content delivery is kept allocating via $y_{aep}=\lfloor c_l/b \rfloor$ until $\sum_{e\in\mathcal{E}^0}\sum_{p\in\mathcal{P}_{ae}}y_{aep}=\lambda_{a}$, i.e. assigning traffic flow depends on the transmission link capacity; otherwise, $y_{aep}=0$.

\begin{algorithm}[!htbp]
	\caption{Greedy Simulated Annealing Caching (GSAC)}
	\label{alg:GSAC}
	\KwData{Variables in Table \ref{tab:notations}; 
	Simulated Annealing Initial Temperature $T_0$;
	Stopping Temperature $T_{end}$;
	Annealing Parameter $\gamma$; 
	Markov-chain Length $L$}
	\KwResult{Cache allocation $x_{ne}$ and path delivery $y_{naep}$}
	Initialize $\mathcal{N'}=\{n|\sum_{a\in\mathcal{A}}\lambda_{na}>0\}$ and sort descendingly\;
	Generate $\{p|p\in\mathcal{P}_{ae}\}$ the k-shortest path for each $(a,e)$ pair via Yen's Algorithm\;
	\ForEach(\tcp*[f]{Simulated Annealing}){$n\in\mathcal{N'}$}{
		Initialize $x_e$ randomly $\{x_e\in\{0,1\}|s_n\!\leq\!w_e,\forall e\!\in\!\mathcal{E}^0\}$, $T_m\!\leftarrow\!T_0$, $j\!\leftarrow\!1$\;
		Determine $y_{aep}$ through \eqref{fml:det_y1} or \eqref{fml:det_y2}\tcp*{for $(P1)$ or $(P2)$}
		Set best solution $x^*,y^*\leftarrow x_{e},y_{aep}$\;
		\While{$T_m\geq T_{end}$}{
			\While{$j\leq L$}{
				Generate $x'_e$ via flipping one bit in $x_e$ randomly\;
				Determine $y'_{aep}$ through \eqref{fml:det_y1} or \eqref{fml:det_y2}\;
				\eIf{$\delta=E(x'_e,y'_{aep})-E(x_e,y_{aep})<0$}{
					Update $x_e,y_{aep}\leftarrow x'_e,y'_{aep}$\;
					\If{$E(x'_e,y'_{aep})-E(x^*,y^*)<0$}{
					Update $x^*,y^*\leftarrow x'_e,y'_{aep}$\;
					}
				}{
					Update $x_e,y_{aep}\leftarrow x'_e,y'_{aep}$ with probability $\exp(-\delta/T_m)$\;
				}
			$j\leftarrow j+1$\;
			}
		$T_m\leftarrow \gamma\times T_m$\;	
		}
	$x_{ne},y_{naep}\leftarrow x^*,y^*$\;
	Update $w_e,c_l$\;	
	}
\end{algorithm}
 
The details of the proposed GSAC framework is depicted in Algorithm \ref{alg:GSAC}. There are four hyperparameters involved as explained in the sequel. The $L$ parameter controls the internal loop from line $8$ to $20$, whilst parameters $T_0$, $T_{end}$ and $\gamma$ manage the external loop from line $7$ to $21$. Moreover, the current temperature $T_m$ influences the acceptance probability, i.e. the exploration of the searching space. 
In line $2$, we construct the k-shortest path via Yen's Algorithm, whose the time complexity is $O(k|\mathcal{E}|^3)$ and symbol $|~|$ represents set cardinality. In this paper, we consider the 5-shortest paths, i.e. $k=5$, and we require the path for each AR $a$. Consequently, the time complexity of line $2$ becomes $O(|\mathcal{A}||\mathcal{E}|^3)$.
For each external iteration of simulated annealing, the temperature is reduced by $\gamma$ from $T_0$ to $T_{end}$. Suppose $I$ is the number of external loops, then there is $\gamma^I\cdot T_0\leq T_{end}$ when terminating. In most cases, the cooldown factor $\gamma$ is between $0$ and $1$. Therefore, we have $I=\lceil\log_{\gamma}(T_{end}/T_0)\rceil$. It is clear that the number of internal loops is $L$. In each internal loop, the worst case is enumerating the combination of $a$, $e$ and $p$ when determining decision variable $y_{aep}$, and, hence the upper bound time complexity for internal loop is $O(|\mathcal{A}||\mathcal{E}||\mathcal{P}|)$. Note that we need to assign caching location for each content $n\in\mathcal{N}'$. In the end, the time complexity of Algorithm \ref{alg:GSAC} becomes $O(IL\cdot|\mathcal{A}||\mathcal{E}||\mathcal{P}||\mathcal{N}'|+|\mathcal{A}||\mathcal{E}|^3)$.  

The configuration of hyperparameters depends on the scope of problem. In this paper, after several rounds of grid search and by considering the trade-off between time complexity and performance, the best configuration is recorded as follows: $T_0=10^3$, $T_{end}=10^{-3}$, $\gamma=0.8$ and $L=200$.
\section{Numerical Investigations}
\label{sec:numerical}

\subsection{Simulation Setting}
In this section, we compare the caching allocation derived from the set of proposed schemes against some well known previously proposed techniques such as the  Random Caching and the Greedy Caching schemes, which have been introduced in~\cite{ioannou2015taxonomy} and~\cite{wang2019proactive} respectively. The proposed set of techniques include the ILP optimization model (which is viewed as offering the optimal solution in this paper, hence in the figures to follow we use the term Optimal to represent the results of the ILP model) and the GSAC technique. Random Caching can be deemed as a technique that can be easily adopted and used by operators where hosting nodes are chosen randomly for the requested content. The Greedy Caching allocates content requests to the nearest node; hence requests are packed to those closest cache nodes if  there is sufficient caching space. 

The performance is evaluated mainly via two key metrics: the energy gain and cache-hit ratio. As for the energy gain, we use the so-called No-caching approach as the benchmark, which, in essence, means that we do not host any content in caching nodes (i.e. $x'_{ne}\equiv 0, \forall n\in\mathcal{N},e\in\mathcal{E}^0$) and each flow request should be transferred to a core network content server (i.e. for the multicast mode $y'_{na\{0\}p}\equiv 1$; for the unicast mode $y'_{na\{0\}p}\equiv\lambda_a$, $\forall n\in\mathcal{N}, a\in\mathcal{A}, \exists p\in\mathcal{P}_{a\{0\}}$). In that case, the gain is calculated by
$$Gain=\frac{E(x'_{ne},y'_{naep})}{E(x_{ne},y_{naep})},$$
where $E(x'_{ne},y'_{naep})$ is the energy cost by applying No-caching strategy, and $E(x_{ne},y_{naep})$ represents the energy consumption by using estimated algorithms, i.e. the Optimal, the GSAC, the Greedy Caching and the Random Caching. As a result, a larger energy gain indicates a better energy-saving performance.
Furthermore, the cache-hit ratio is defined as the percentage of content requests served by caching nodes. 

The experiments are performed in a 10-node 16-link network model, which captures a nominal tree-like configuration of a mobile network. Without loss of generality, we assume that all links have equal capacity and all caching nodes have identical memory space, but the actual value varies to investigate the trade-off between bandwidth and storage (caching memory capacity). Additionally, we assume that the content requests follow the typically used 80/20 rule, i.e. the majority of predictions focus on $20\%$ of the content categories~\cite{paschos2019cache}. The caching allocation is based on such prediction and the impact of predicted accuracy is also discussed. 

All results presented hereafter are averaged over one hundred network scenarios (i.e., Monte Carlo simulations) and based on those we analyse the performance of different schemes. To put the computational times in context we note that all simulations performed using MATLAB R2021a running on top of a 64-bit Ubuntu 20.04.2 LTS environment, whilst the machine is equipped with an Intel Core i7-6700HQ CPU with 32GB RAM. The simulation parameters assumed in the investigations are presented and summarized in Table \ref{tab:sim}. Unless stated otherwise, the value of $w_e$, $c_l$, $\rho_{kn}$, and $\sum_a\lambda_{na}$ are set as the default.

\begin{table}[!htbp]
	\centering
	\caption{Simulation Parameters\cite{zahed2019content}}
	\label{tab:sim}
	\small{\singlespacing
		\begin{tabular}{lcc}
			\toprule
			\textbf{Parameter}&\textbf{Default Value}&\textbf{Range}\\
			\midrule
			caching power coefficient ($\alpha$)& $2.5\times10^{-9}$ W/b&-\\
			transmitting energy coefficient ($\beta$)& $4\times10^{-8}$ J/b&-\\
			content categories ($|\mathcal{N}|$)& 100&-\\
			content size ($s_n$)& - & [10,100] MB\\
			content transmitting bandwidth ($b_n$)& - & [10,100] Mbps\\
			available caching space ($w_e$)& 0.5 GB & [0.25,1] GB\\
			available link capacity ($c_l$)& 0.5 Gbps & [0.25,1] Gbps\\
			preference prediction accuracy ($\rho_{kn}$)& $100\%$ & [0\%,100\%]\\
			number of requests ($\sum_a\lambda_{na}$) & $100$ &[20,160]\\
			time slot duration ($T$)& 10 s&-\\
			\bottomrule
		\end{tabular}
	}
\end{table}

\subsection{Impact of the Number of Requests}

\begin{figure}[htbp]
	\centering
	
	\begin{subfigure}{\figSizeOne\textwidth}
		\centering
		\includegraphics[width=\textwidth]{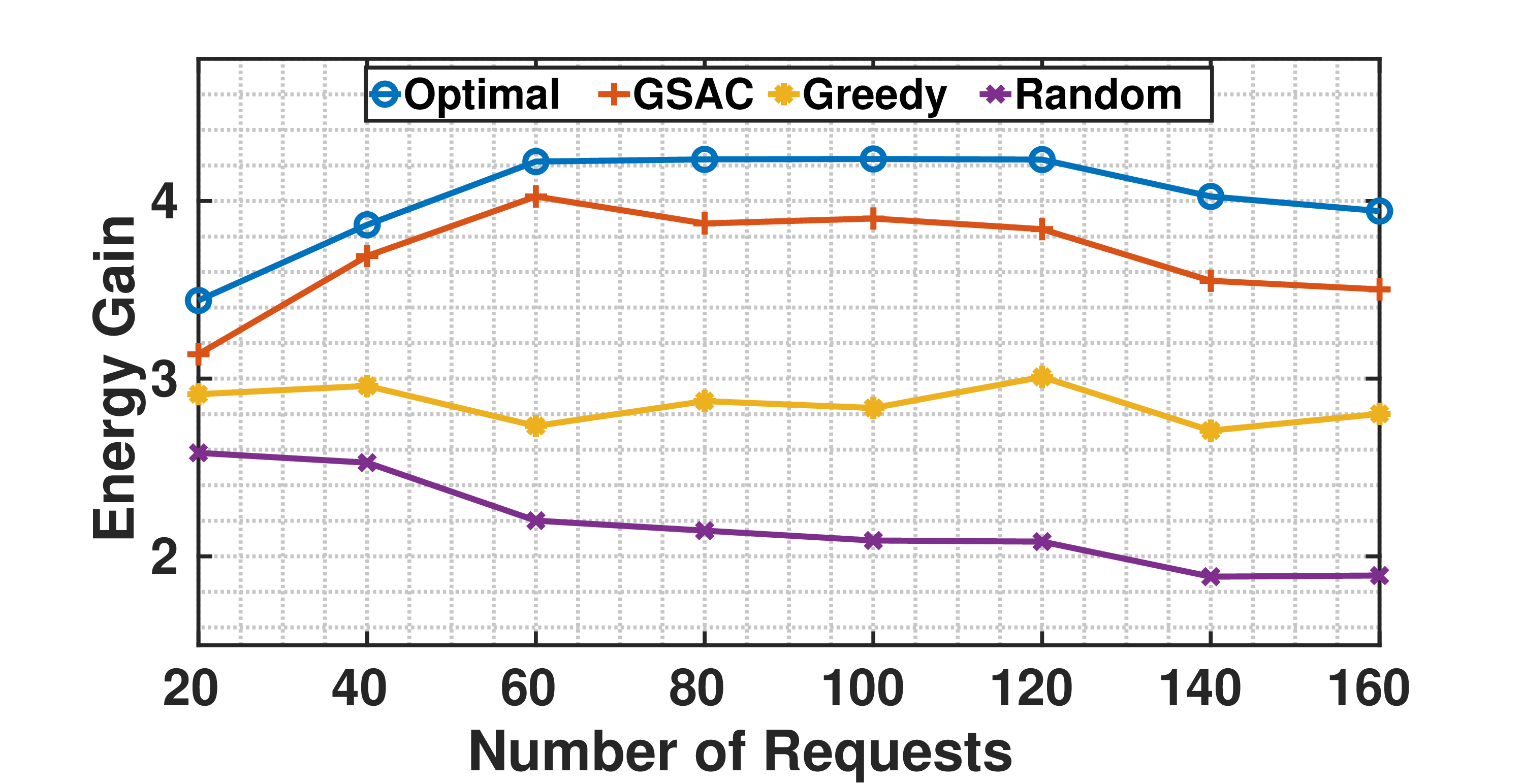}
		\caption{Energy Gain on P1}
		\label{fig:EG_P1}
	\end{subfigure}
	\begin{subfigure}{\figSizeOne\textwidth}
		\centering
		\includegraphics[width=\textwidth]{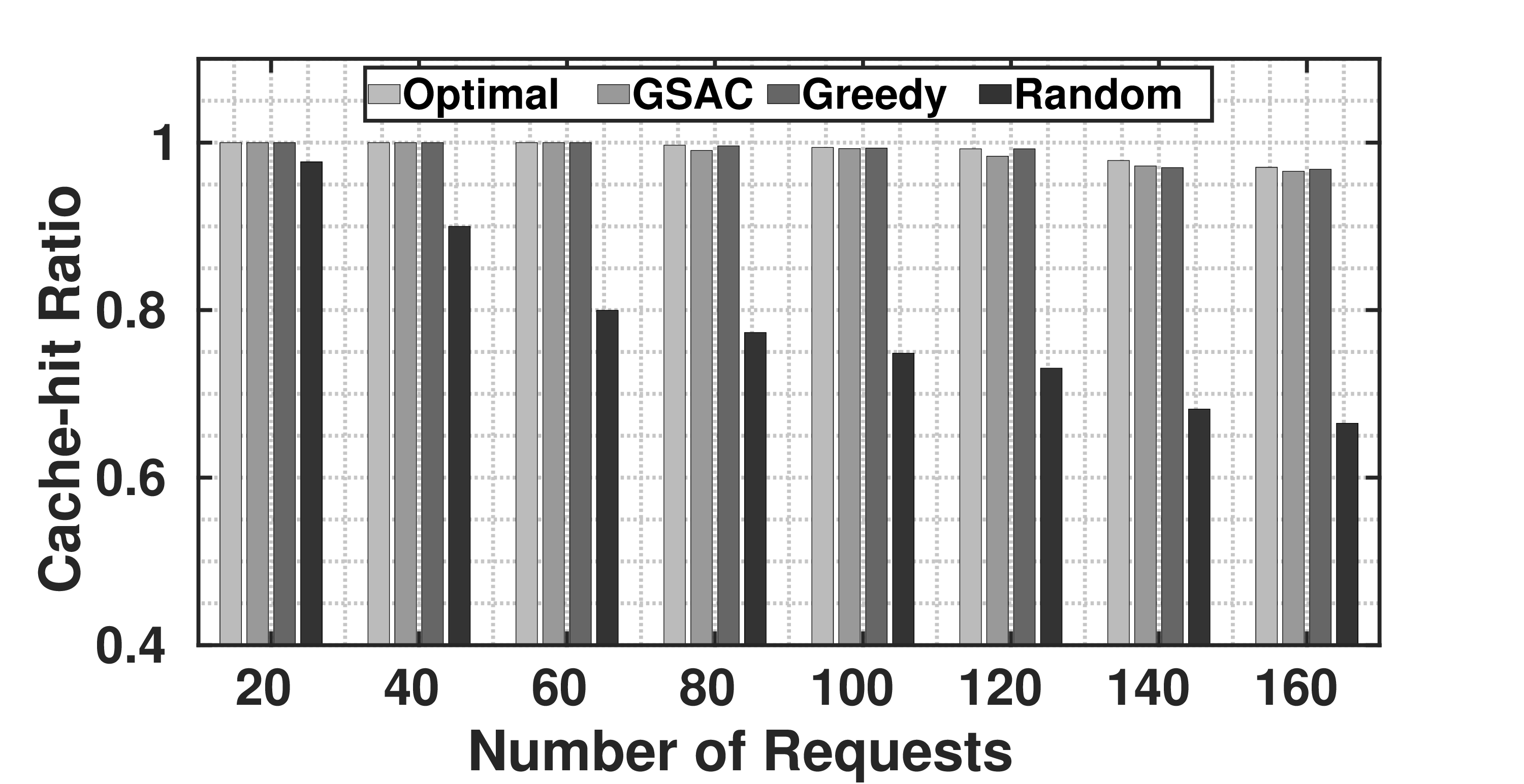}
		\caption{Cache-hit Ratio on P1}
		\label{fig:CR_P1}
	\end{subfigure}
	
	\begin{subfigure}{\figSizeOne\textwidth}
		\centering
		\includegraphics[width=\textwidth]{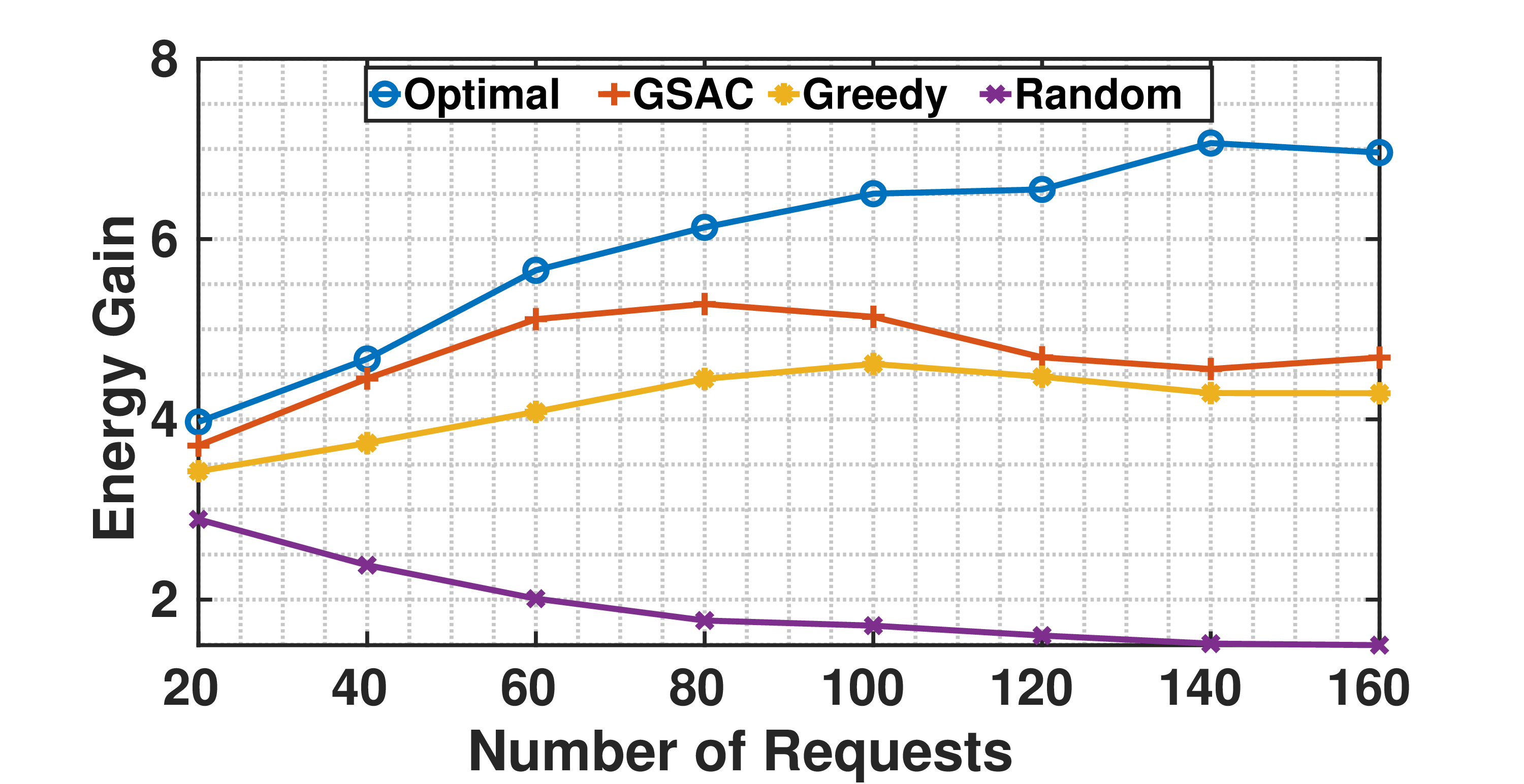}
		\caption{Energy Gain on P2}
		\label{fig:EG_P2}
	\end{subfigure}
	\begin{subfigure}{\figSizeOne\textwidth}
		\centering
		\includegraphics[width=\textwidth]{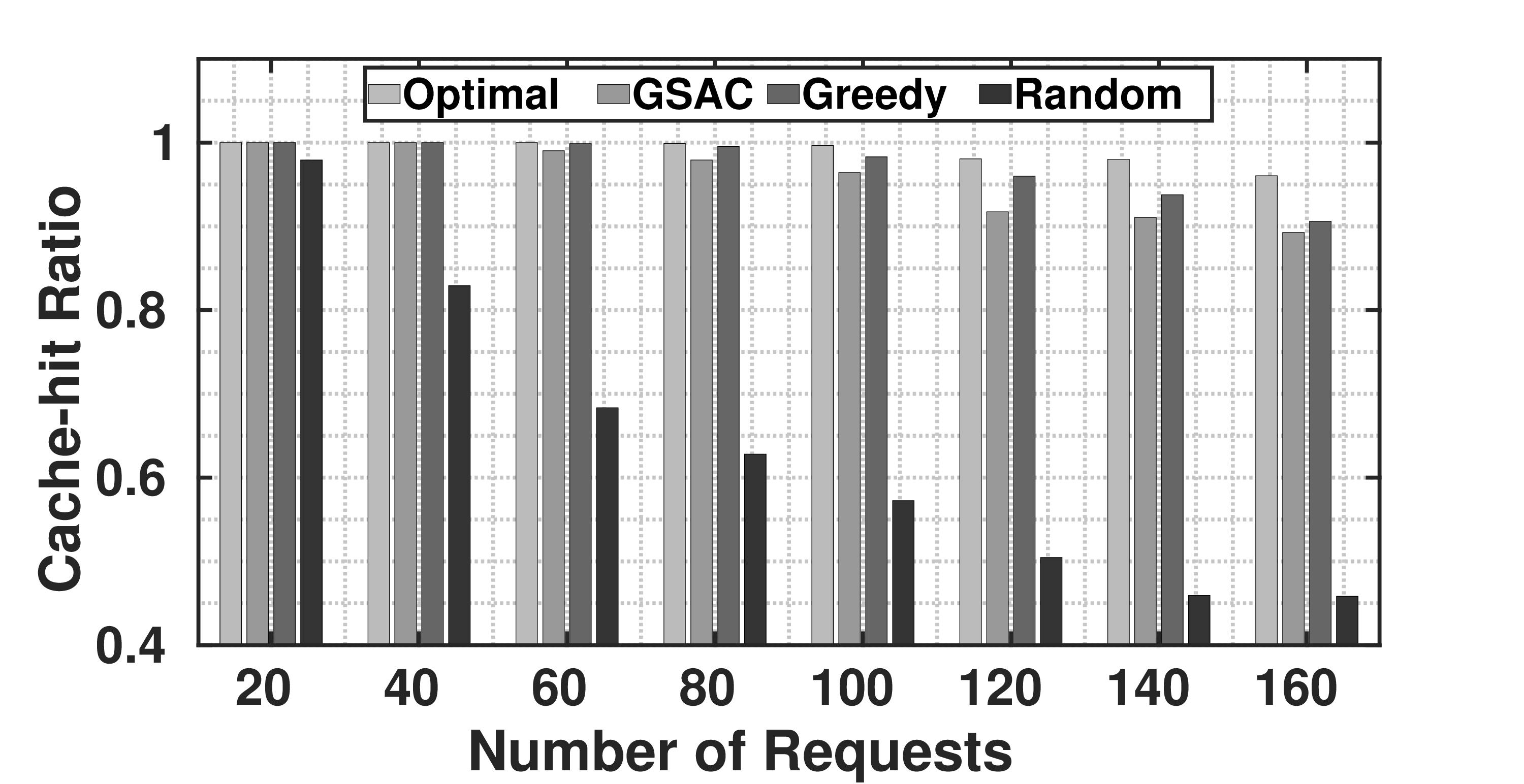}
		\caption{Cache-hit Ratio on P2}
		\label{fig:CR_P2}
	\end{subfigure}
	
	\begin{subfigure}{\figSizeOne\textwidth}
		\centering
		\includegraphics[width=\textwidth]{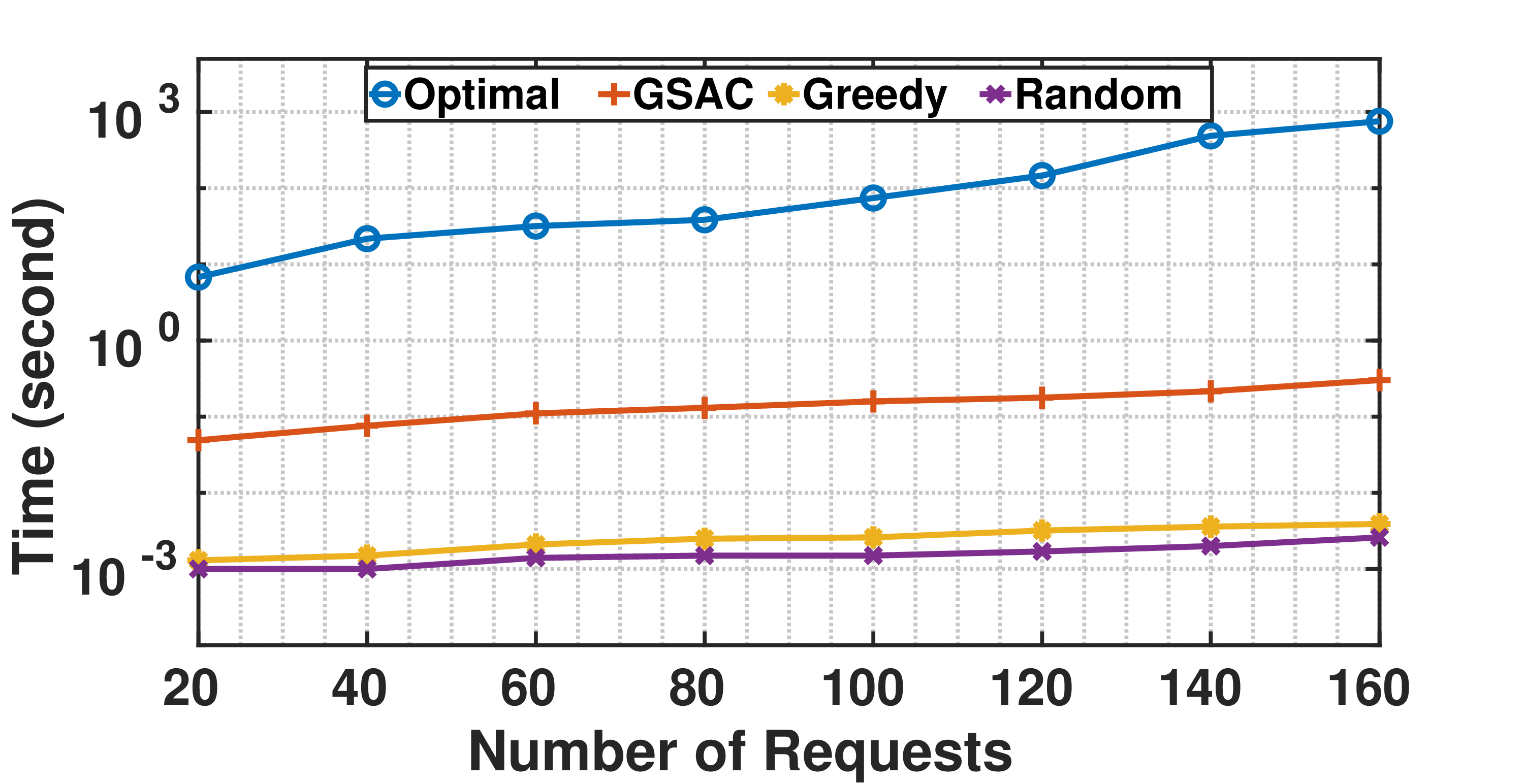}
		\caption{Time Complexity on P1}
		\label{fig:Time_P1}
	\end{subfigure}
	\begin{subfigure}{\figSizeOne\textwidth}
		\centering
		\includegraphics[width=\textwidth]{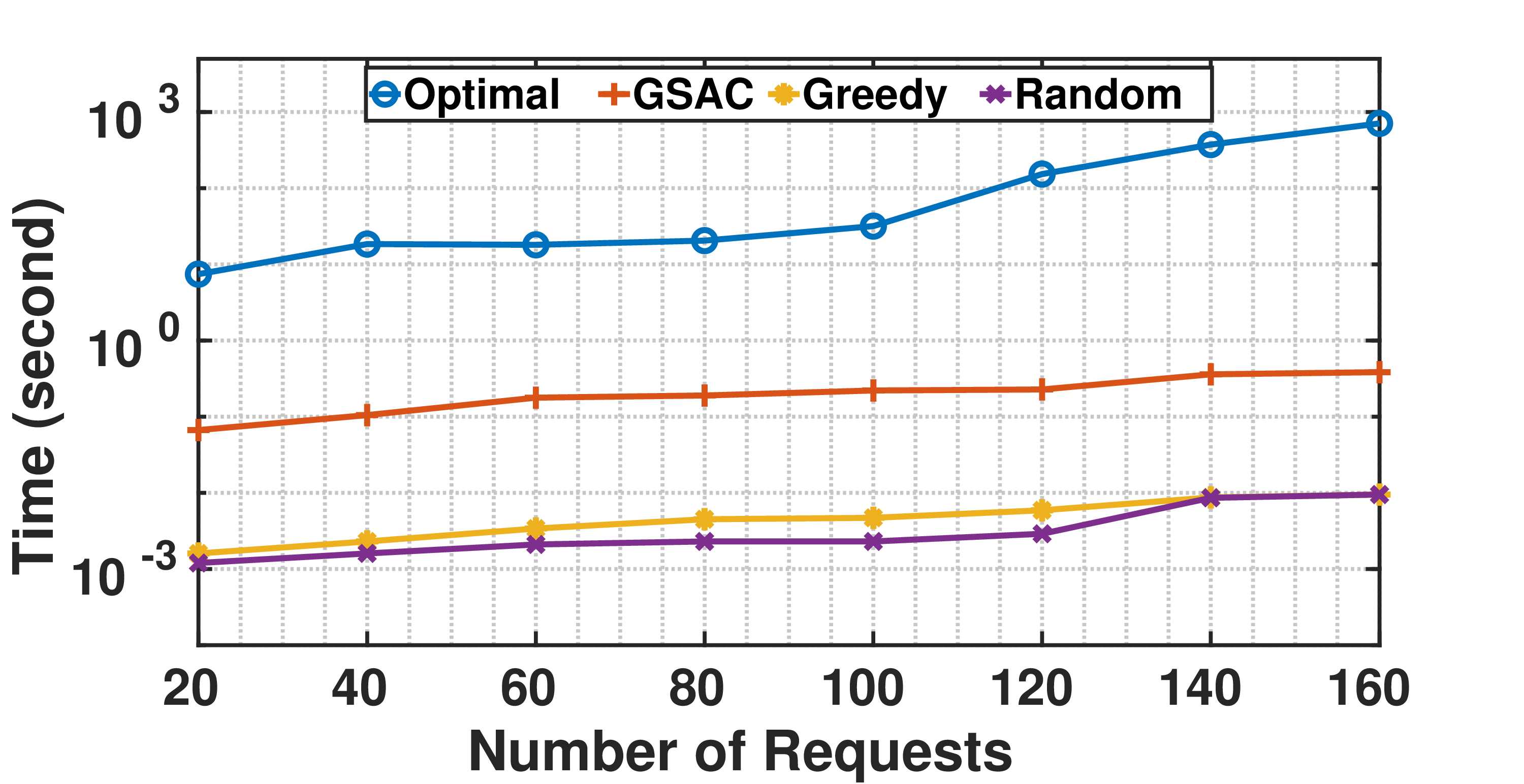}
		\caption{Time Complexity on P2}
		\label{fig:Time_P2}
	\end{subfigure}
	\caption{Algorithms Performance versus the Number of Requests}
	\label{fig:per_requests}
\end{figure}

We first study how the number of content requests affect the performance of proposed schemes, in which case  we set $\sum_a\lambda_{na}$ to vary from $20$ to $160$. In Figure \ref{fig:per_requests}.(\subref{fig:EG_P1}), both the Optimal and GSAC on $(P1)$ share a climbing-remaining-dropping trend while the Random has a decreasing trend. With the increase in user requests, as expected, the energy consumption for caching service is increasing accordingly, and this trend is independent if we utilize the Optimal scheme, the GSAC, the Greedy, the Random, or the No-caching. However, the rate of increase plays a significant role in the overall energy gains: 1) an increase (i.e., positive slope) represents the No-caching scheme grows at a faster rate than the evaluated algorithm; 2) a plateau (i.e., zero slope) shows that both the No-caching and the evaluated algorithm have similar increasing rates; and 3) a decrease (i.e., negative slope) indicates that the No-caching grows slower than the evaluated algorithm. 

In Figure \ref{fig:per_requests}.(\subref{fig:EG_P1}), when the number of requests ranges from 20 to 60, the network has sufficient resources (i.e. caching space and link bandwidth) to serve end-users, and the Optimal together with GSAC can both significantly outperform the No-caching scheme in reducing energy consumption. The energy gain ranges from $3.44$ to $4.23$ for the Optimal, and $3.13$ to $3.87$ for the GSAC scheme. Then, passing that state, the network becomes saturated when the number of requests ranges between 60 and 120, which reflects a plateau   on the energy gain curve. After 120 requests, more users' flows are redirected to a core network server, hence the performance resembles more to the No-caching scheme. Therefore, the energy gain is reduced to $3.94$ for the Optimal and $3.50$ for GSAC. Regarding Greedy Caching, the energy gain fluctuates between $2.70$ and $3.01$ depending on the number of requests. Since it endeavours to assign the requested content to the nearest node, Greedy Caching provides savings in the transmission energy but pays more cost for hosting the contents. For Random Caching, observe the gradual decline on the energy gain from $2.58$ to $1.88$ in the simulation, which indicates this method cannot deal with the increasing requests in a flexible manner. Finally, it is worth noting that the energy gain in Figure \ref{fig:per_requests}.(\subref{fig:EG_P1}) is above 1, so all evaluated algorithms can save energy compared to the No-caching approach. 

Figure \ref{fig:per_requests}.(\subref{fig:CR_P1}) reveals that there is a continuous decrease in the cache-hit ratio as the number of requests increases. Generally, the Optimal, GSAC and Greedy Caching guarantee that the majority of requests (over $95\%$ in the simulation) can be satisfied on the caching nodes, while nearly a third of the flows are redirected to a core network server by applying Random Caching in the 160-request case. Notably, the ratios of Optimal and GSAC have marked decrease after 120 requests, which match the turning points in Figure \ref{fig:per_requests}.(\subref{fig:EG_P1}). 

A similar pattern still holds on $(P2)$ as shown in Figure \ref{fig:per_requests}.(\subref{fig:EG_P2}) and \ref{fig:per_requests}.(\subref{fig:CR_P2}). However, in $(P2)$ the energy gain of the Optimal keeps rising beyond the 120-request point. The reason is that in $(P2)$ each flow is served independently, even if the requested content is in essence the same. Thus, the No-caching algorithm consumes more energy than in $(P1)$. Although the Optimal energy is also increasing, it is at a slower rate than the No-caching, which reflects a positive slope when the number of requests ranges from 120 to 140. After 120 requests, the performance gap between GSAC and Greedy in Figure \ref{fig:per_requests}.(\subref{fig:EG_P2}) gets smaller. In contrast to the dropping trend in \ref{fig:per_requests}.(\subref{fig:EG_P1}), the lack of more generalized hyperparameters of GSAC should be blamed instead of network resource limitation. Because the Optimal solution keeps increasing which indicates the better allocation exists but GSAC cannot find it. Similar to $(P1)$, the gain of Random Caching keeps decreasing with  increasing number of requests. Compared with Random Caching, the Optimal can save $80\%$ energy in the 140-request case, which is also the maximum attained performance gain in the simulations. 
Interestingly, the actual influence of the different delivery modes on energy gain and cache-hit ratio vary: 
on the one hand, unicast mode $(P2)$ generates more traffic flows which make the energy gain higher than $(P1)$ by comparing Figure \ref{fig:per_requests}.(\subref{fig:EG_P1}) and \ref{fig:per_requests}.(\subref{fig:EG_P2}) vertically; on the other hand, more flows exhaust the network resource rapidly and, thus, the cache-hit ratio in \ref{fig:per_requests}.(\subref{fig:CR_P2}) is less than \ref{fig:per_requests}.(\subref{fig:CR_P1}).  

The running time of $(P1)$ and $(P2)$ are evaluated in Figure \ref{fig:per_requests}.(\subref{fig:Time_P1}) and \ref{fig:per_requests}.(\subref{fig:Time_P2}) respectively. The time for Optimal solution varies from 6.80 seconds to around 13 minutes, while the proposed heuristic algorithm GSAC is less than 400 milliseconds and Greedy Caching as well as Random Caching are less than 10 milliseconds. 

In Figure \ref{fig:per_requests}, the Optimal outperforms the other methods in both $(P1)$ and $(P2)$ problems, albeit with a higher time complexity cost. The heuristic method GSAC is able to provide a competitive result, followed by Greedy and Random. For instance, compared with the Optimal allocation, in $(P1)$ GSAC saves $99.8\%$ running time with the cost of approximate $7.9\%$ performance loss when there are 100 requests. Under the same scenario, the time-saving ratio for Greedy Caching and Random Caching is $99.9\%$ compared with the Optimal solution; but the performance reductions are $33.1\%$ and $50.7\%$ respectively. 


\subsection{Impact of the Available Network Capacity (in terms of bandwidth and storage)}

To understand the trade-off between the bandwidth and caching space, the values of the parameters $w_e$ and $c_l$ are varied from 0.25 to 1 with a sampling interval of 0.25. This setting aims to shed light on the following key question: when the memory space and link capacity are insufficient to support all content requests, to further improve the performance (i.e. energy gain and cache-hit ratio), should we increase the link capacity first or the caching space?  

\begin{figure}[htbp]
	\centering
	\begin{subfigure}{\figSizeTwo\textwidth}
		\centering
		\includegraphics[trim=80mm 0mm 100mm 10mm,clip, width=\textwidth]{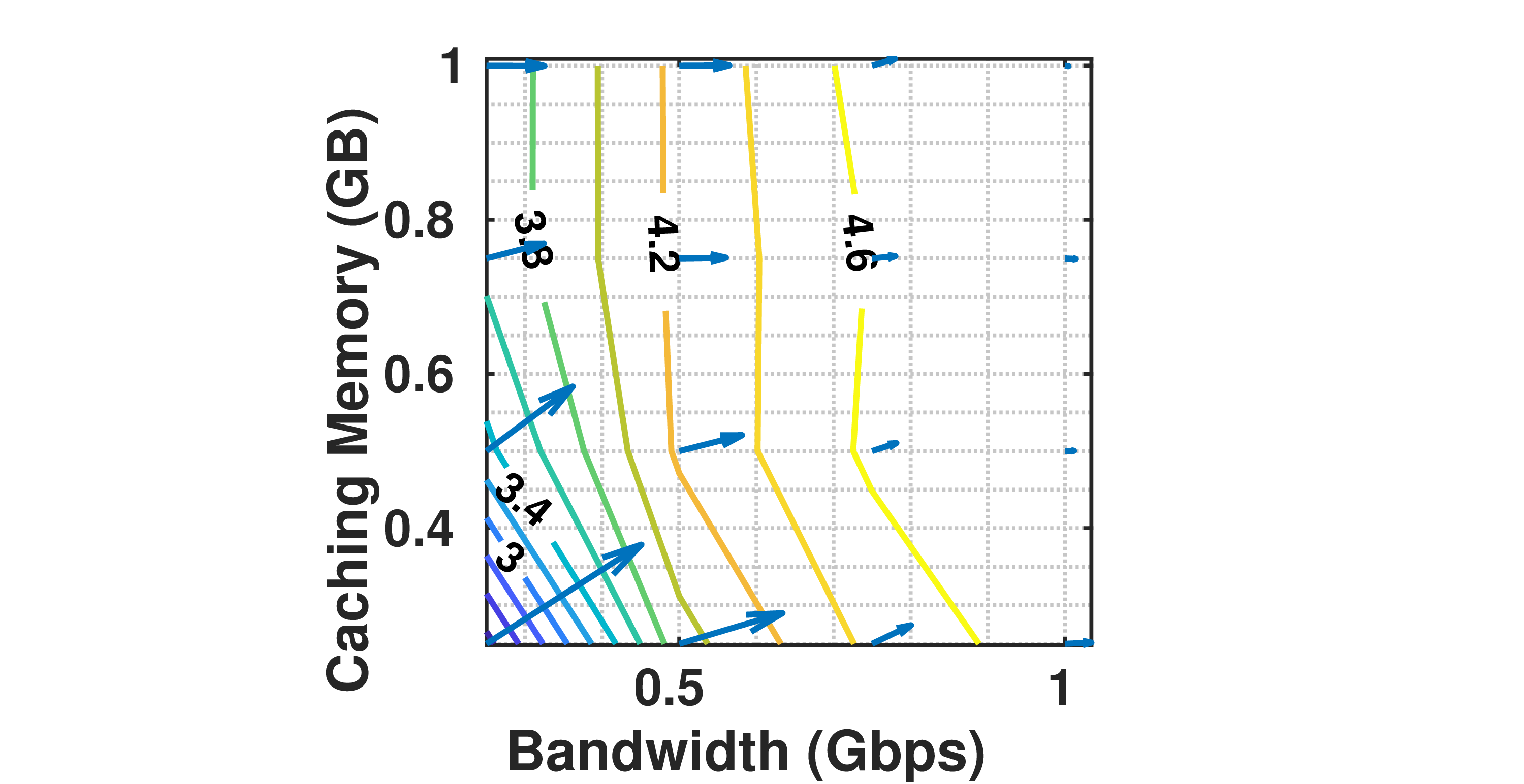}
		\caption{Optimal Energy Gain on P1}
		\label{fig:EG_P1_Optimal}
	\end{subfigure}
	\begin{subfigure}{\figSizeTwo\textwidth}
		\centering
		\includegraphics[trim=80mm 0mm 100mm 10mm,clip, width=\textwidth]{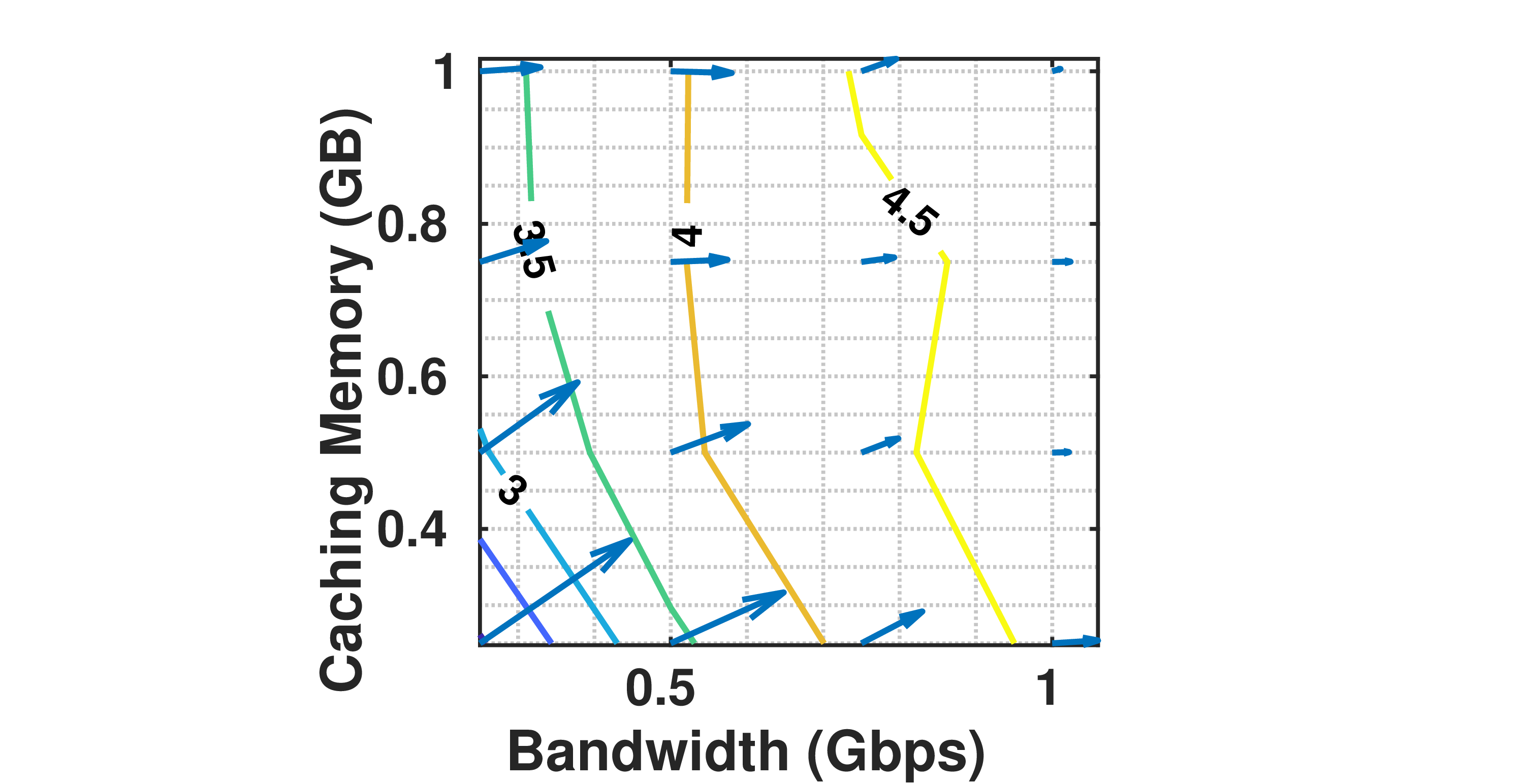}
		\caption{GSAC Energy Gain on P1}
		\label{fig:EG_P1_GSAC}
	\end{subfigure}
	\begin{subfigure}{\figSizeTwo\textwidth}
		\centering
		\includegraphics[trim=80mm 0mm 100mm 10mm,clip, width=\textwidth]{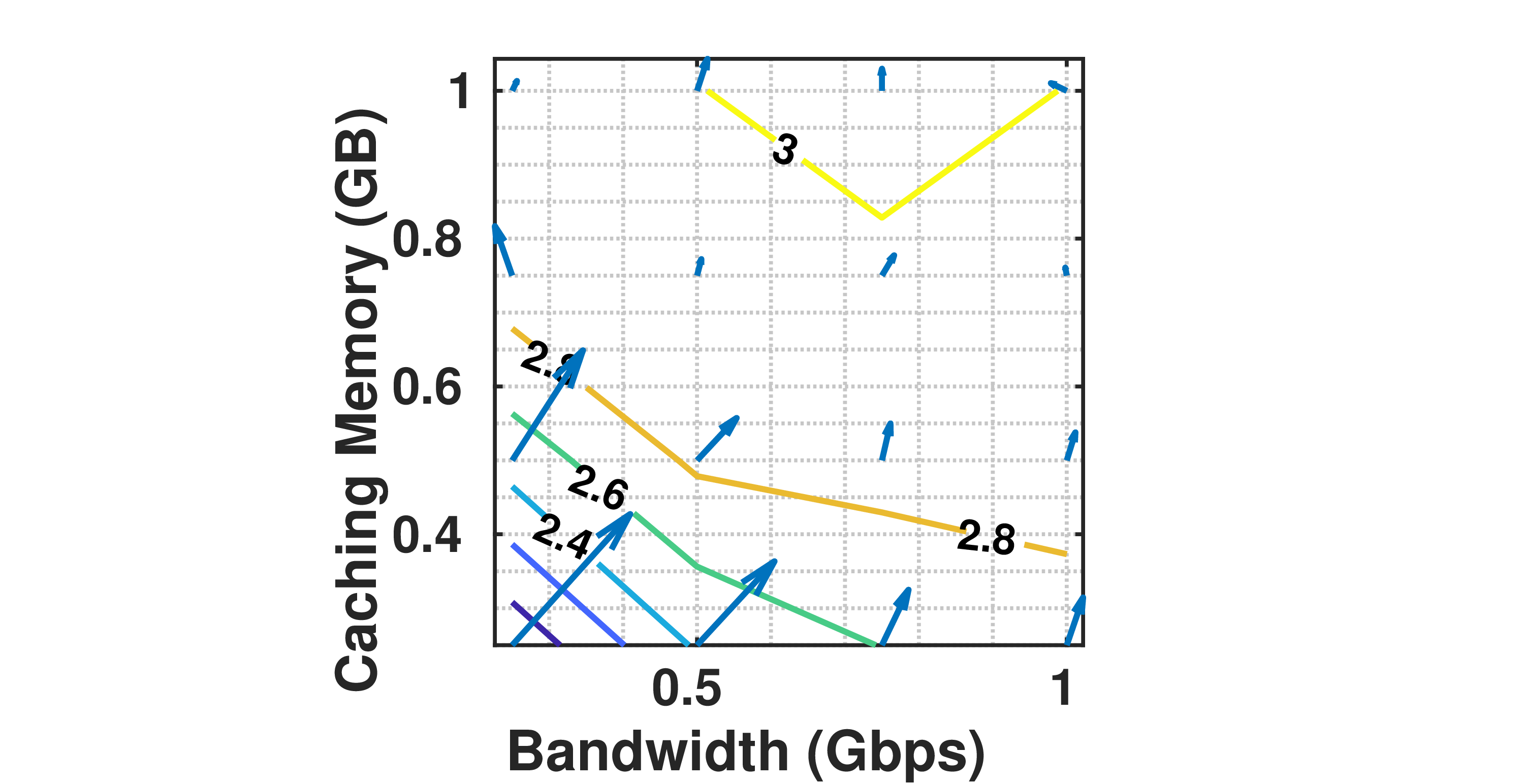}
		\caption{Greedy Energy Gain on P1}
		\label{fig:EG_P1_Greedy}
	\end{subfigure}
	\begin{subfigure}{\figSizeTwo\textwidth}
		\centering
		\includegraphics[trim=80mm 0mm 100mm 10mm,clip, width=\textwidth]{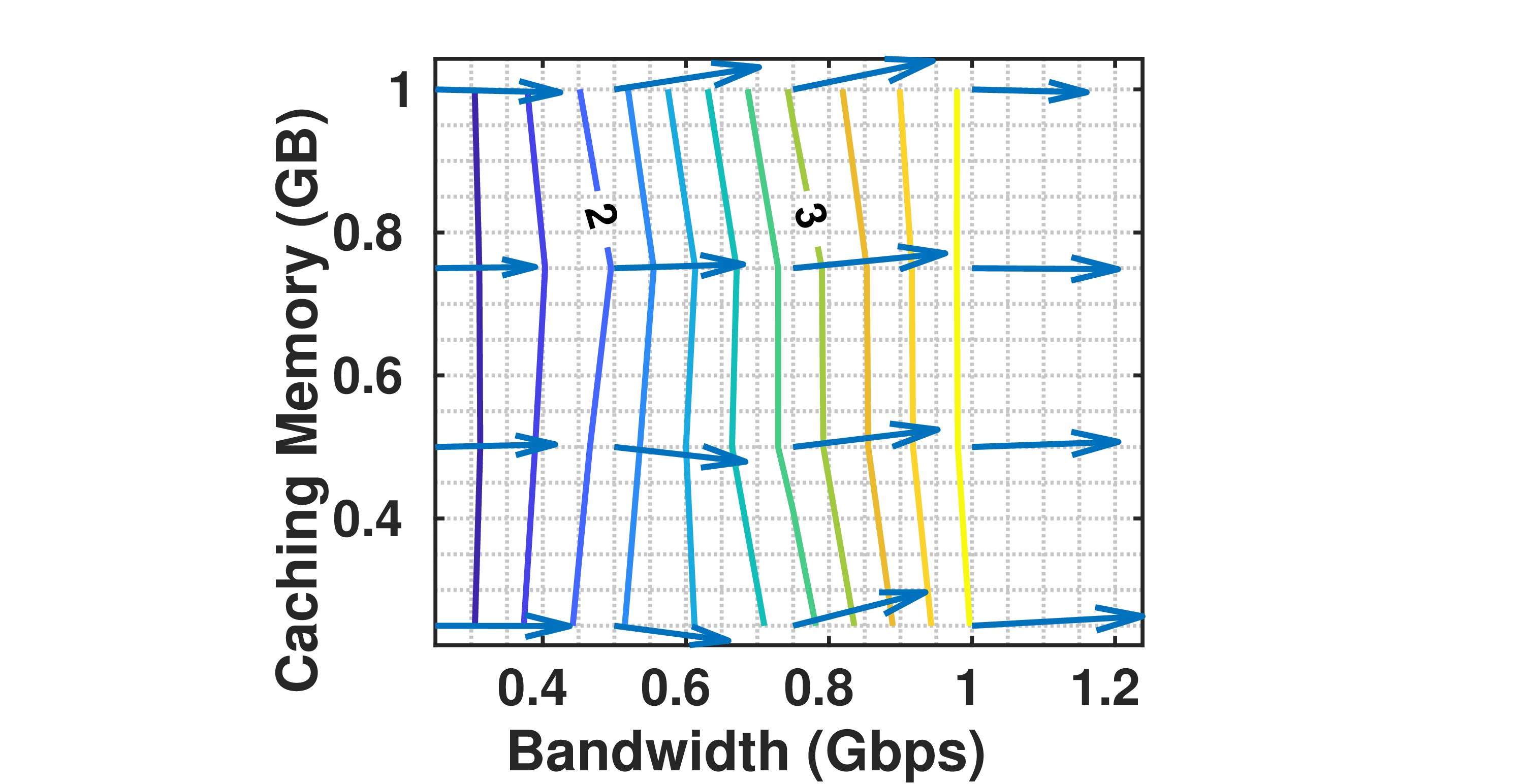}
		\caption{Random Energy Gain on P1}
		\label{fig:EG_P1_Random}
	\end{subfigure}

	\begin{subfigure}{\figSizeTwo\textwidth}
		\centering
		\includegraphics[trim=80mm 0mm 100mm 10mm,clip, width=\textwidth]{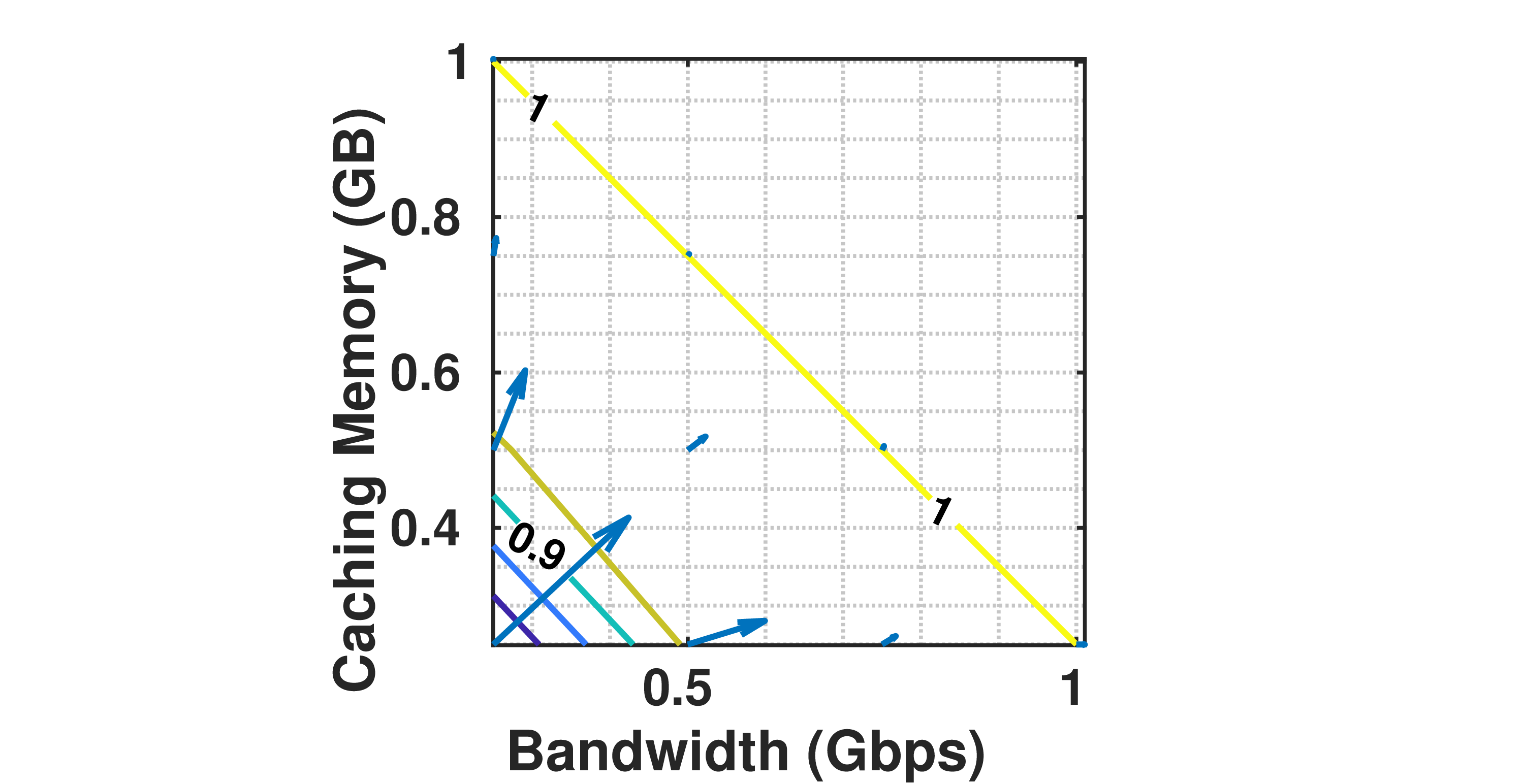}
		\caption{Optimal Cache-hit Ratio on P1}
		\label{fig:CH_P1_Optimal}
	\end{subfigure}
	\begin{subfigure}{\figSizeTwo\textwidth}
		\centering
		\includegraphics[trim=80mm 0mm 100mm 10mm,clip, width=\textwidth]{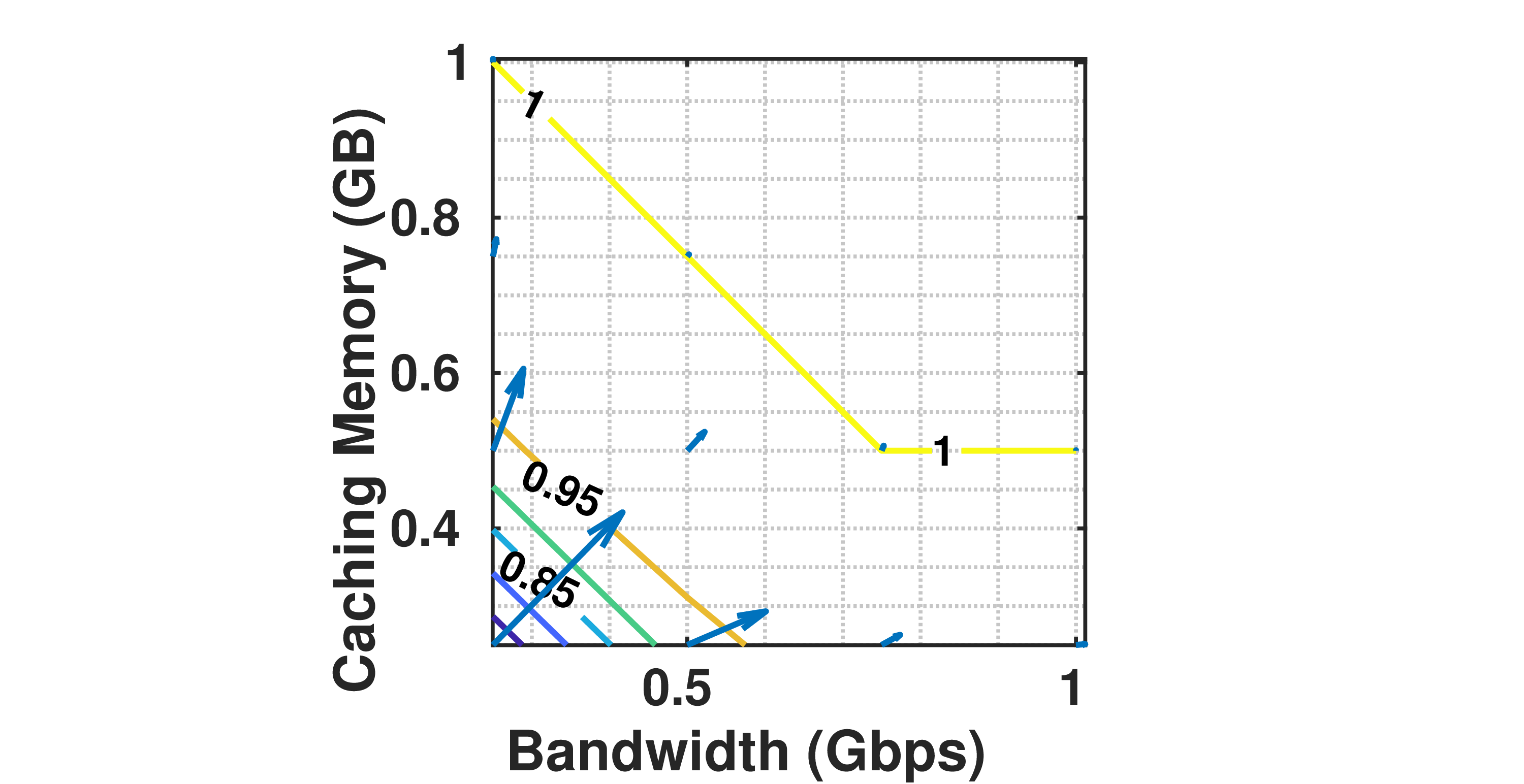}
		\caption{GSAC Cache-hit Ratio on P1}
		\label{fig:CH_P1_GSAC}
	\end{subfigure}
	\begin{subfigure}{\figSizeTwo\textwidth}
		\centering
		\includegraphics[trim=80mm 0mm 100mm 10mm,clip, width=\textwidth]{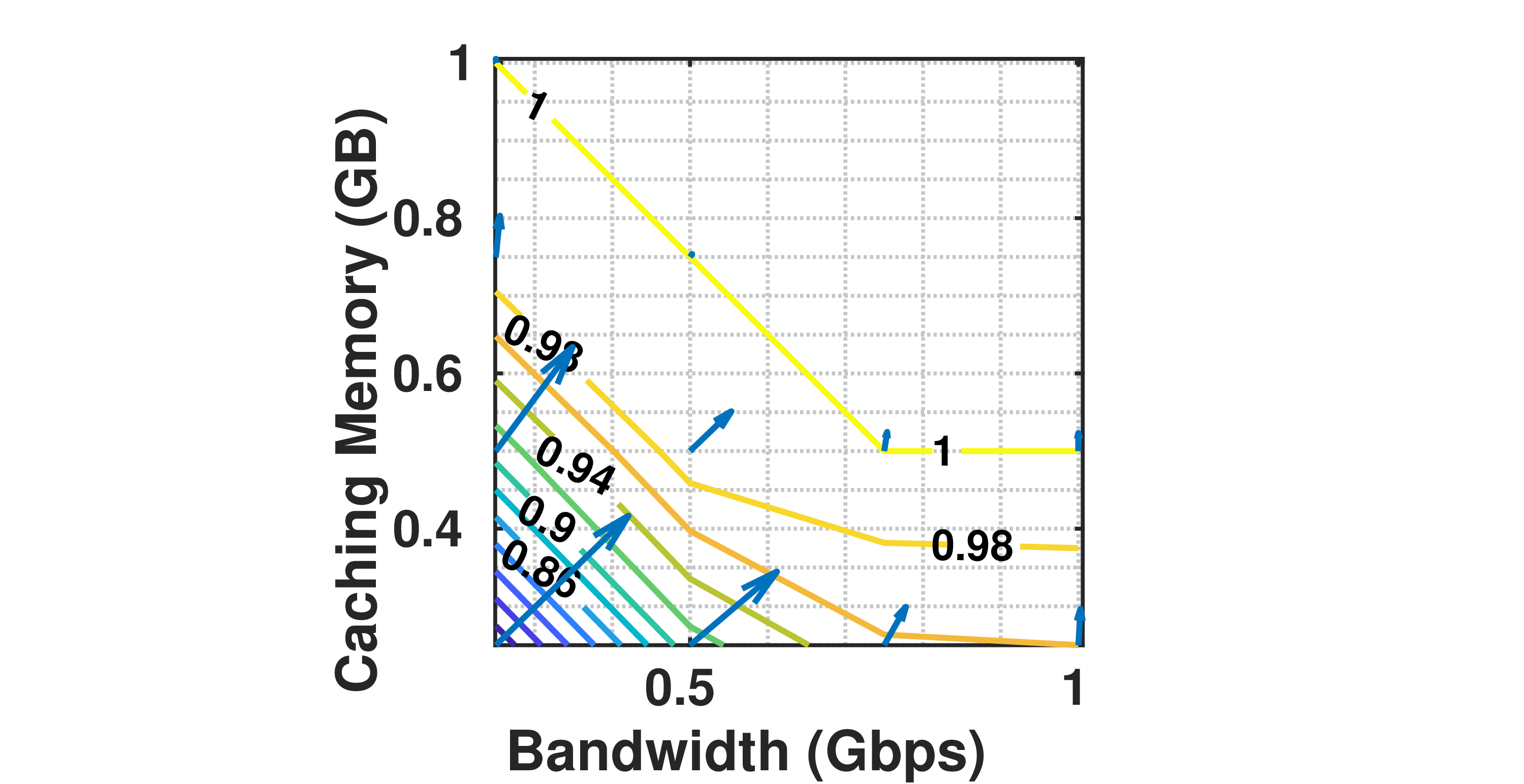}
		\caption{Greedy Cache-hit Ratio on P1}
		\label{fig:CH_P1_Greedy}
	\end{subfigure}
	\begin{subfigure}{\figSizeTwo\textwidth}
		\centering
		\includegraphics[trim=80mm 0mm 100mm 10mm,clip, width=\textwidth]{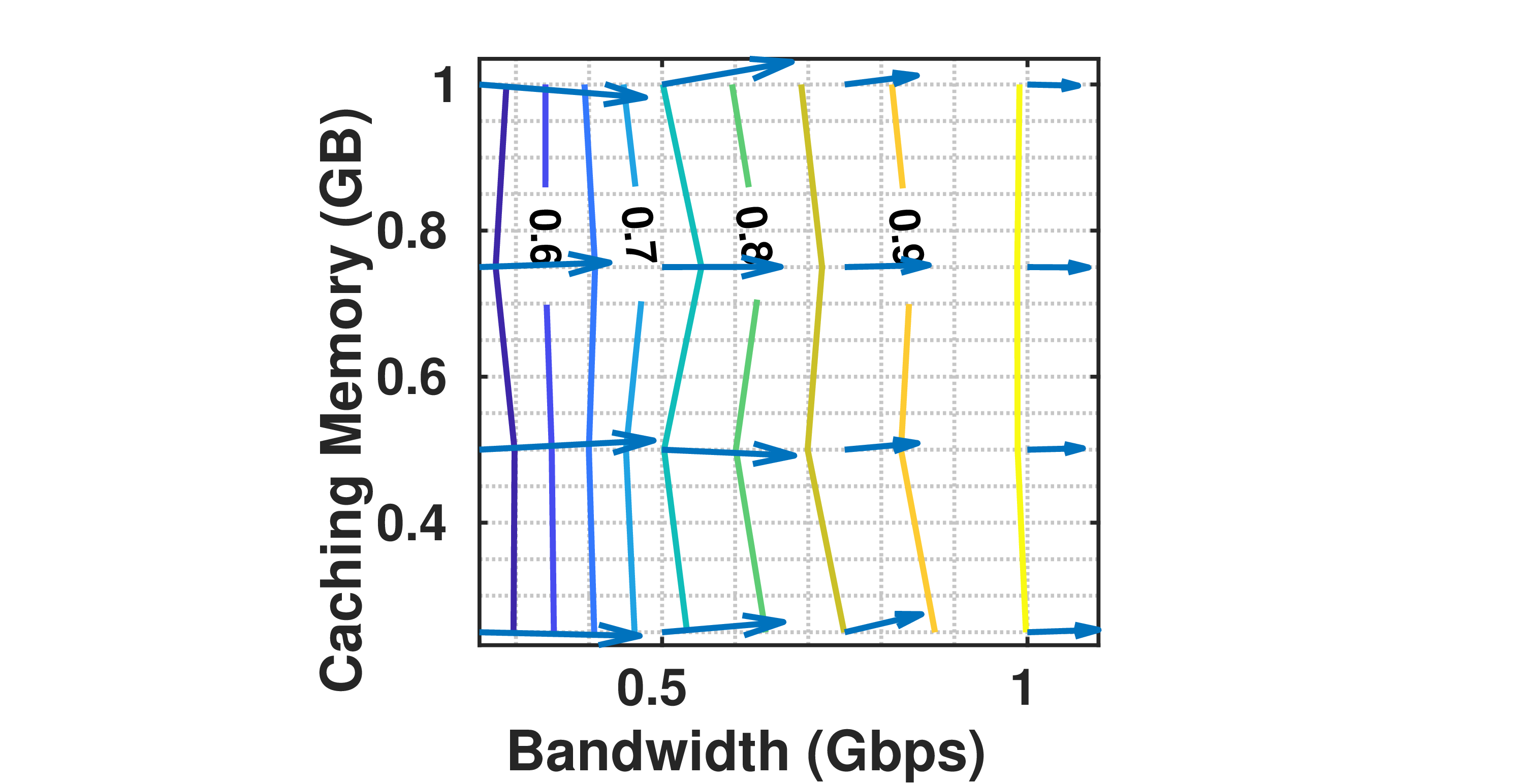}
		\caption{Random Cache-hit Ratio on P1}
		\label{fig:CH_P1_Random}
	\end{subfigure}

	\begin{subfigure}{\figSizeTwo\textwidth}
		\centering
		\includegraphics[trim=80mm 0mm 100mm 10mm,clip, width=\textwidth]{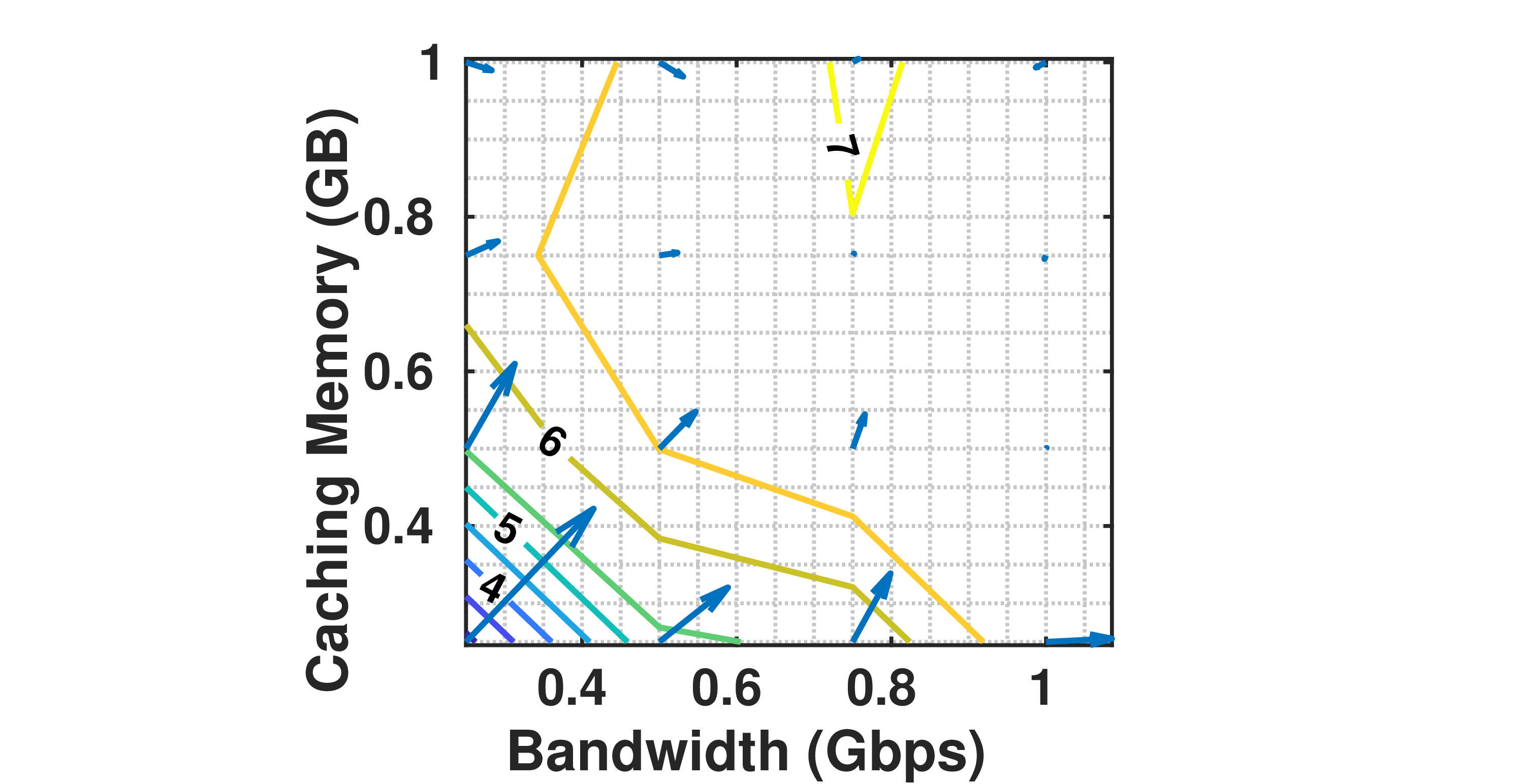}
		\caption{Optimal Energy Gain on P2}
		\label{fig:EG_P2_Optimal}
	\end{subfigure}
	\begin{subfigure}{\figSizeTwo\textwidth}
		\centering
		\includegraphics[trim=80mm 0mm 100mm 10mm,clip, width=\textwidth]{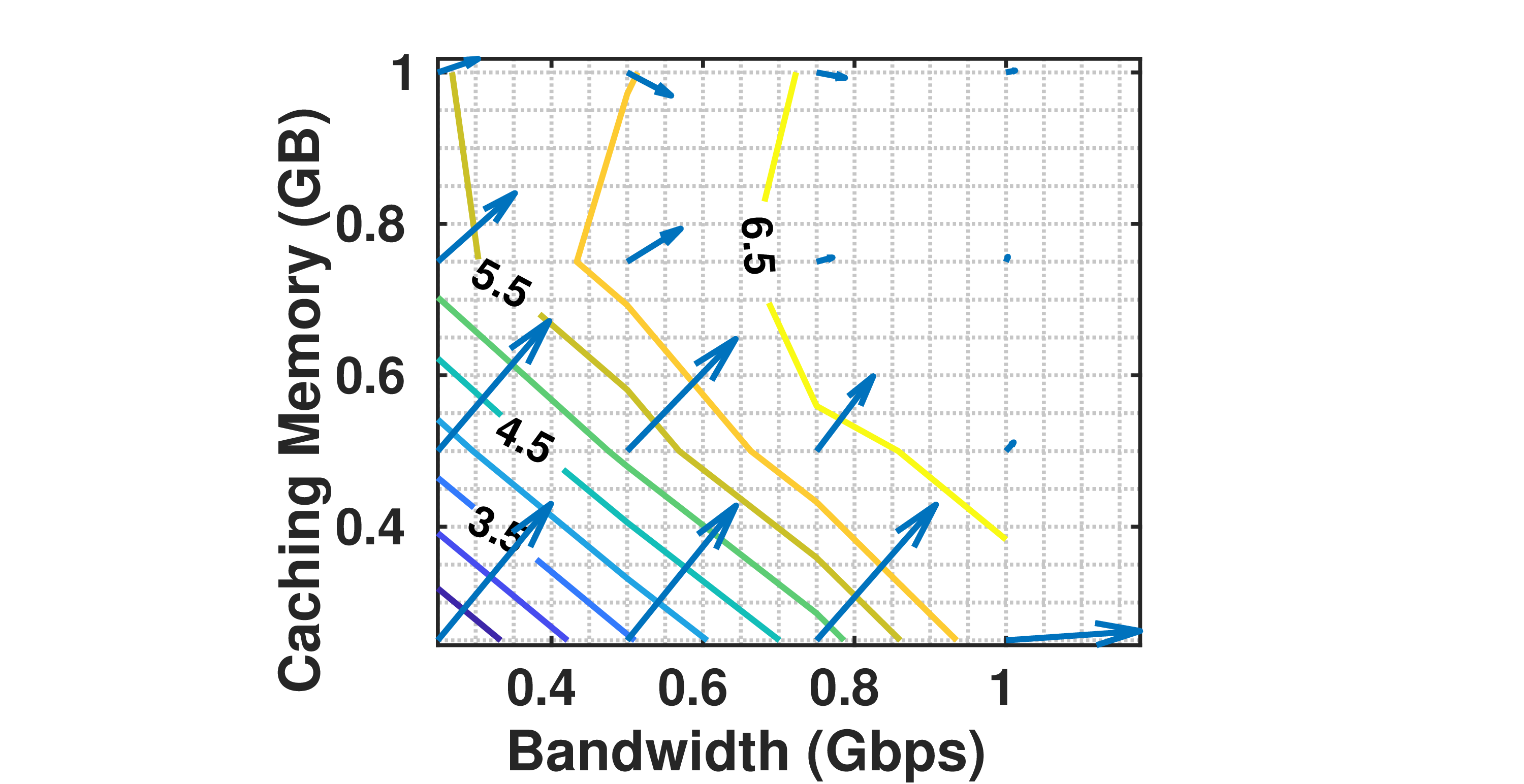}
		\caption{GSAC Energy Gain on P2}
		\label{fig:EG_P2_GSAC}
	\end{subfigure}
	\begin{subfigure}{\figSizeTwo\textwidth}
		\centering
		\includegraphics[trim=80mm 0mm 100mm 10mm,clip, width=\textwidth]{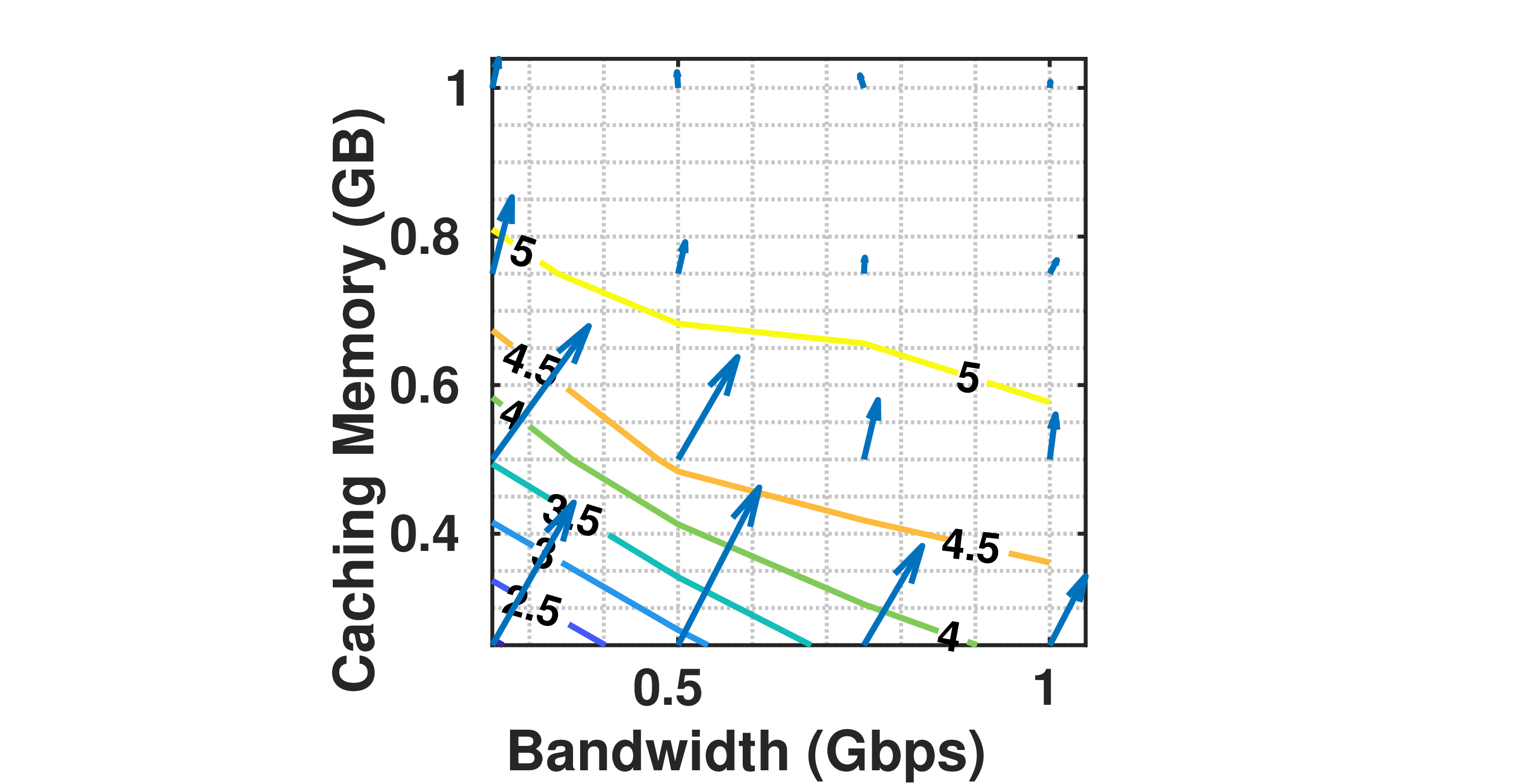}
		\caption{Greedy Energy Gain on P2}
		\label{fig:EG_P2_Greedy}
	\end{subfigure}
	\begin{subfigure}{\figSizeTwo\textwidth}
		\centering
		\includegraphics[trim=80mm 0mm 100mm 10mm,clip, width=\textwidth]{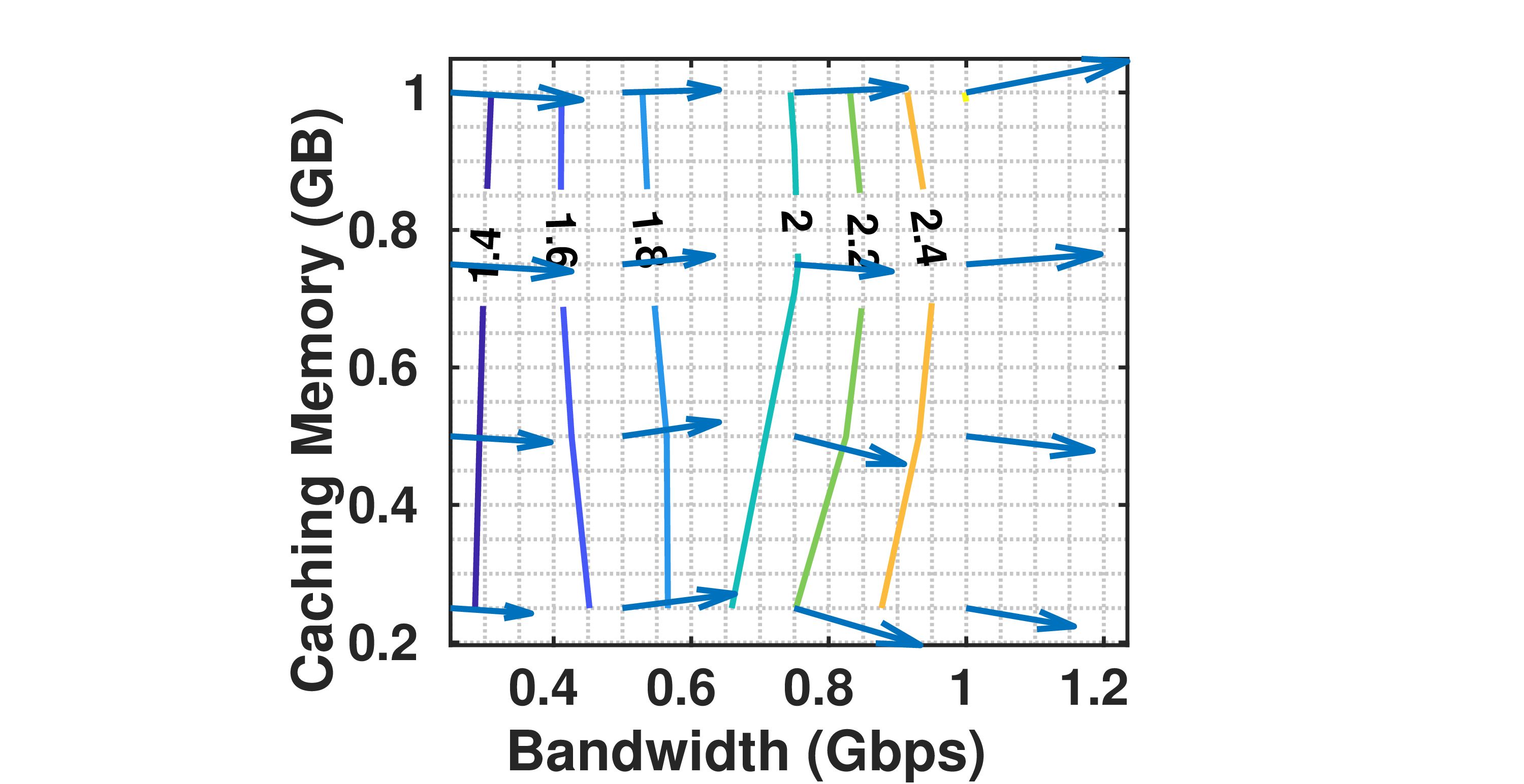}
		\caption{Random Energy Gain on P2}
		\label{fig:EG_P2_Random}
	\end{subfigure}

	\begin{subfigure}{\figSizeTwo\textwidth}
		\centering
		\includegraphics[trim=80mm 0mm 100mm 10mm,clip, width=\textwidth]{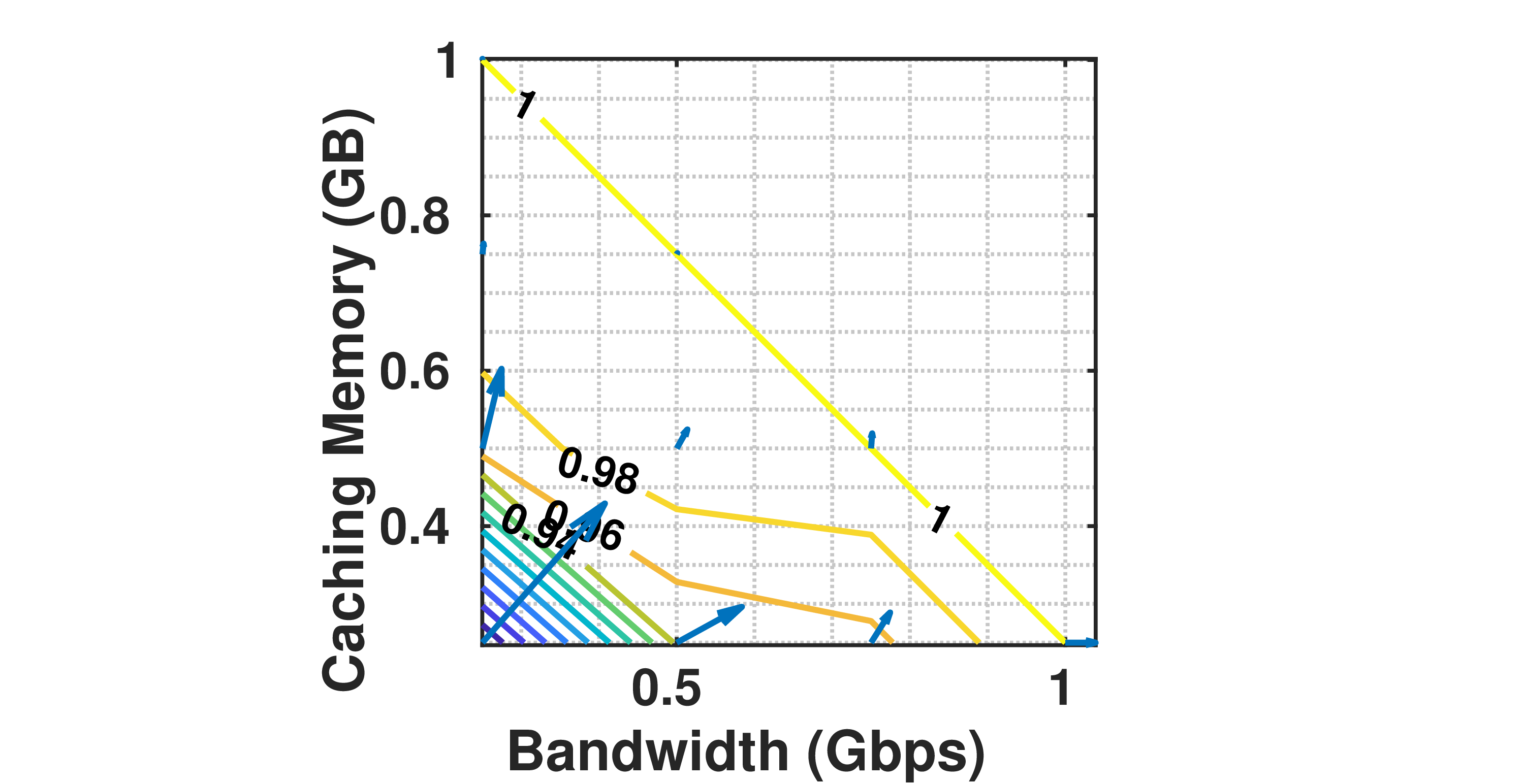}
		\caption{Optimal Cache-hit Ratio on P2}
		\label{fig:CH_P2_Optimal}
	\end{subfigure}
	\begin{subfigure}{\figSizeTwo\textwidth}
		\centering
		\includegraphics[trim=80mm 0mm 100mm 10mm,clip, width=\textwidth]{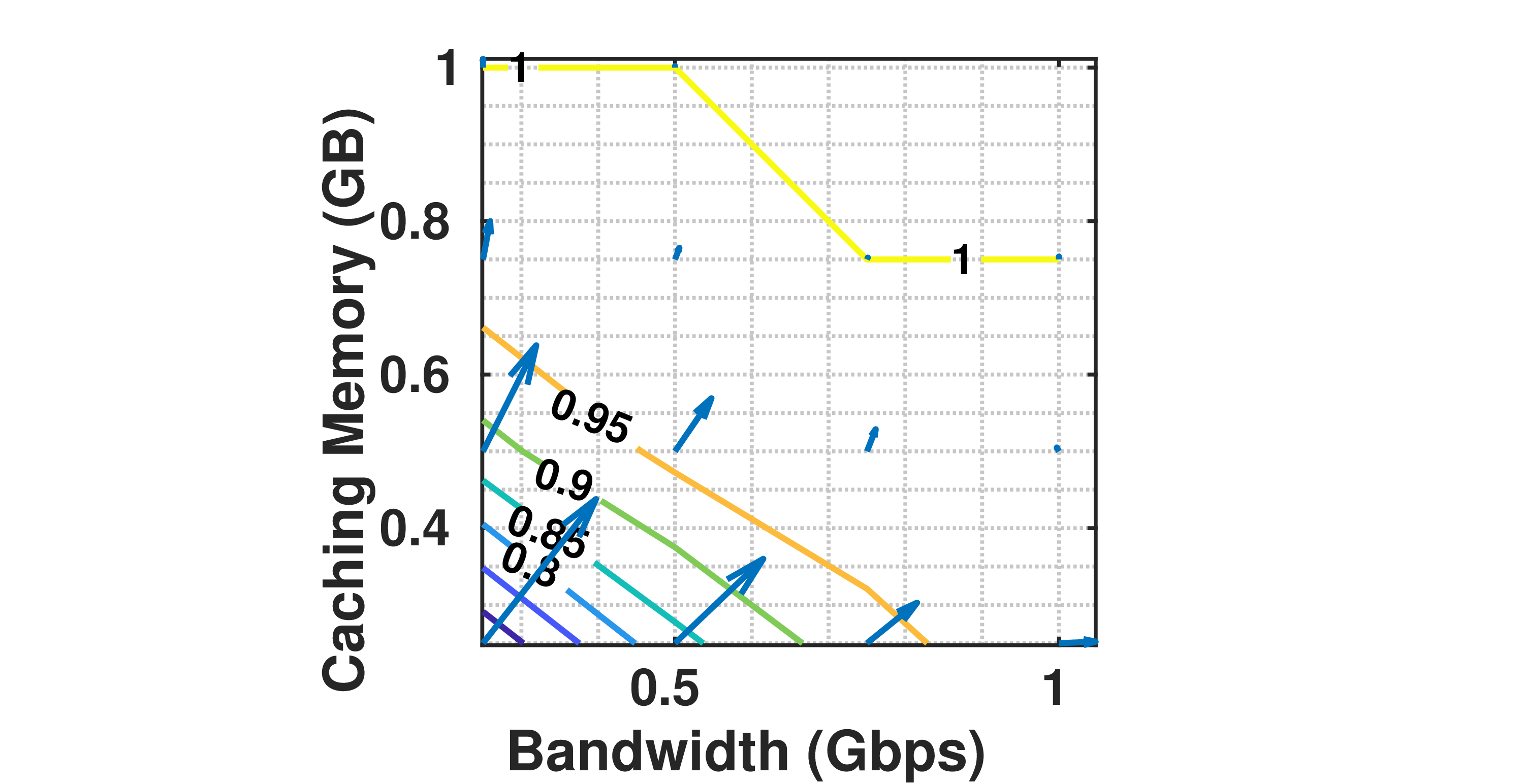}
		\caption{GSAC Cache-hit Ratio on P2}
		\label{fig:CH_P2_GSAC}
	\end{subfigure}
	\begin{subfigure}{\figSizeTwo\textwidth}
		\centering
		\includegraphics[trim=80mm 0mm 100mm 10mm,clip, width=\textwidth]{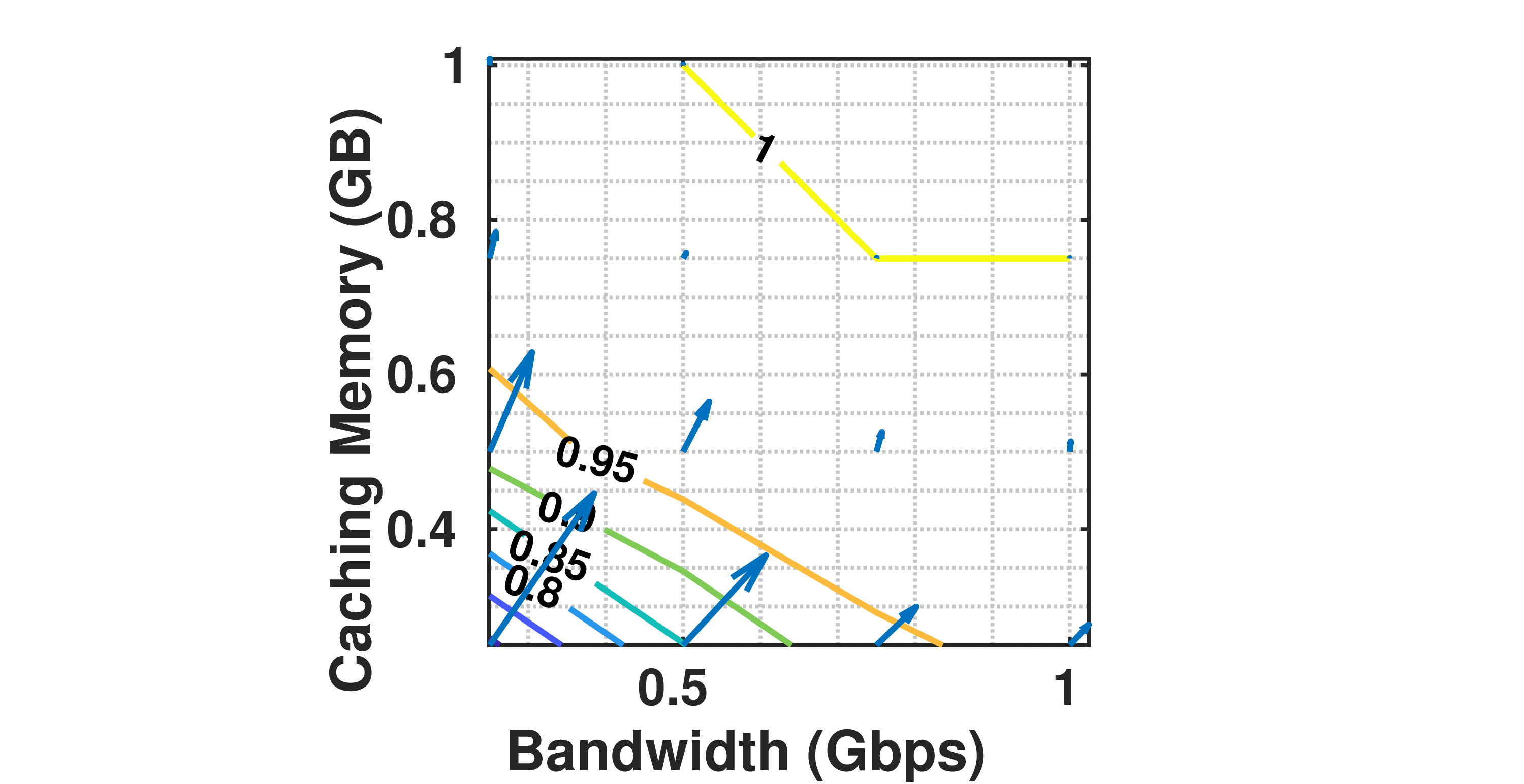}
		\caption{Greedy Cache-hit Ratio on P2}
		\label{fig:CH_P2_Greedy}
	\end{subfigure}
	\begin{subfigure}{\figSizeTwo\textwidth}
		\centering
		\includegraphics[trim=80mm 0mm 100mm 10mm,clip, width=\textwidth]{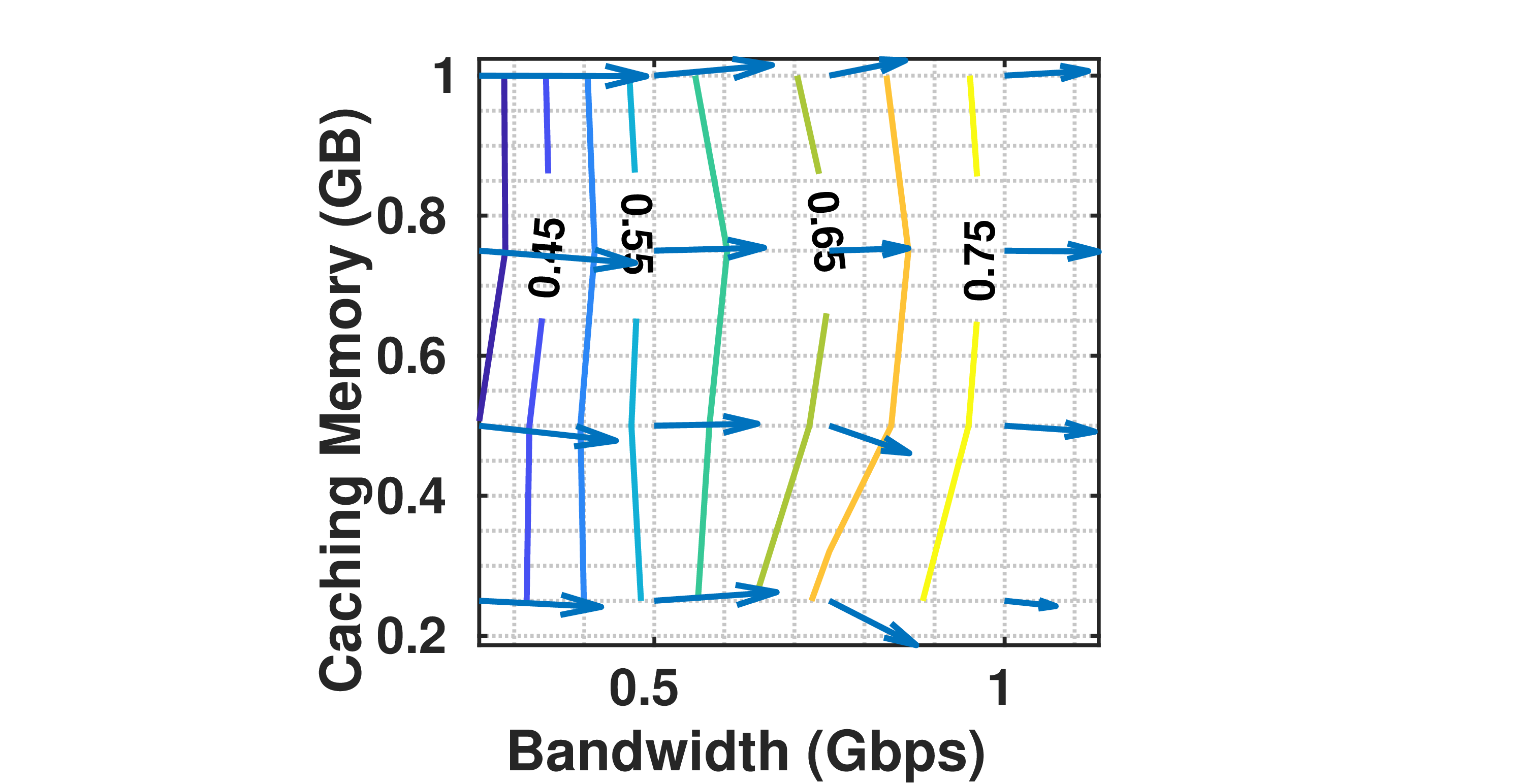}
		\caption{Random Cache-hit Ratio on P2}
		\label{fig:CH_P2_Random}
	\end{subfigure}

	\caption{The Trade-off between Bandwidth and Caching Space}
	\label{fig:per_capacity}
\end{figure}

In Figure \ref{fig:per_capacity} we test the performance of the different algorithms over $(P1)$ and $(P2)$. The contour lines with labels represent the energy gain or cache-hit ratio accordingly. Moreover, the blue arrows depict the gradient, which can be viewed as the directionality towards optimized parameterization. In Figure \ref{fig:per_capacity}.(\subref{fig:EG_P1_Optimal}), when both the bandwidth and memory are less than 0.5, either increasing the caching space or expanding the link capacity results in an energy gain growth. However, once the memory space is larger than 0.5 GB, the benefit from expanding link bandwidth is greater than memory, as shown in Figure \ref{fig:per_capacity}.(\subref{fig:EG_P1_Optimal}) that the arrow points towards the bandwidth increasing direction. To some extent, enlarging the caching memory implies more energy consumption for hosting content. A similar trend holds on Figure \ref{fig:per_capacity}.(\subref{fig:EG_P1_GSAC}), 0.5 GB works as a watershed: before 0.5 GB both bandwidth and caching memory make contributions to the energy gain; after 0.5 GB the bandwidth dominates the performance improvement. Regarding Greedy Caching in Figure \ref{fig:per_capacity}.(\subref{fig:EG_P1_Greedy}), when the bandwidth is above 0.5 Gbps, enlarging the caching space can bring more benefits compared with link capacity increasing, as the arrow points towards the vertical direction. The main reason for its insensitivity on link capacity is that when the memory space is sufficient, Greedy Caching allocates the content replicates in the nearest node, which is the access router in most of the cases. Therefore, each request can be served locally and as such eliminate  bandwidth resource usage. When it comes to $(P2)$, the tendencies of optimal \ref{fig:per_capacity}.(\subref{fig:EG_P2_Optimal}), GSAC \ref{fig:per_capacity}.(\subref{fig:EG_P2_GSAC}) and Greedy \ref{fig:per_capacity}.(\subref{fig:EG_P2_Greedy}) are similar with $(P1)$. What is striking in Figure \ref{fig:per_capacity}.(\subref{fig:EG_P1_Random}) is that all arrows point towards the bandwidth increasing direction, which indicates that expanding link capacity has a distinct effect on the energy gain improvement while the caching space seems of not significantly contributing. Additionally, from Figure \ref{fig:per_capacity}.(\subref{fig:CH_P1_Random}), \ref{fig:per_capacity}.(\subref{fig:EG_P2_Random}), and \ref{fig:per_capacity}.(\subref{fig:CH_P2_Random}), the performance of Random Caching is very sensitive to the bandwidth change but sluggish to the caching memory. For the other cases of cache-hit ratio (Figure \ref{fig:per_capacity}.(\subref{fig:CH_P1_Optimal})$\sim$(\subref{fig:CH_P1_Greedy}), \ref{fig:per_capacity}.(\subref{fig:CH_P2_Optimal})$\sim$(\subref{fig:CH_P2_Greedy})), both caching space and link bandwidth play an equally important role in the performance improvement, as the slope of arrows keeps close to 1. 

Thus, based on the aforementioned analysis the answer on how to prioritize expanding of the caching memory first or the link capacity can be categorized into the following three cases: 1) for the Optimal and GSAC algorithms, both bandwidth and memory expanding lead to better cache-hit ratios. However, raising the link capacity can bring more energy gains, especially when the memory space exceeds 0.5 GB on $(P1)$ or 0.75 GB on $(P2)$; 2) regarding Greedy Caching, the cache-hit ratio is affected by both bandwidth and memory space. Nonetheless, the caching memory space plays a dominant factor in energy gains for capacities larger than 0.5 Gbps; 3) if the Random Caching is utilized then the best network strategy is to prioritize towards bandwidth increase instead of caching space to increase energy reduction benefits.  

\subsection{Impact of the Network Topology}

The influence of different network topologies is investigated in Table \ref{tab:per_topo}. In addition to the nominal tree-like network used in the previous simulations which is marked as original, we also employ a strict-tree structure,   in which only one path exists between any two nodes on the network, and a full mesh topology that provides the maximum diversity in terms of routes between nodes in the network. It is expected that as the degree of a node in the network increases, both the energy gain and cache-hit ratio of the evaluated algorithms can be further improved since providing more direct connections is equivalent to expanding the available link bandwidth, especially for $(P2)$. 
It is also clearly illustrated that in the case of a full mesh network topology, even the use of Random Caching can be considered as a suitable method for energy-saving purposes. Observe however that Greedy Caching is under-performing because the caching energy dominates total energy consumption while its advantage on saving transmission energy becomes less obvious, particularly in a full mesh network topology.

\begin{table}[htbp]
	\centering
	\caption{Algorithms Performance versus Network Topology}
	\label{tab:per_topo}
	\small{\singlespacing{
	\begin{tabular}{cccrccrc}
		\toprule
		\multicolumn{2}{c}{} & \multicolumn{3}{c}{P1} & \multicolumn{3}{c}{P2} \\ 
		\midrule 
		\multicolumn{2}{c}{Topology} & Tree & Original & Full Mesh & Tree & Original & Full Mesh \\ 
		\midrule
		\multicolumn{2}{c}{\# Nodes} & 10 & 10 & 10 & 10 & 10 & 10 \\ 
		\midrule 
		\multicolumn{2}{c}{\# Links} & 9 & 16 & 45 & 9 & 16 & 45 \\ 
		\midrule
		\multirow{4}{*}{\begin{tabular}[c]{@{}c@{}}Energy\\ Gain\end{tabular}} & Optimal & 3.77 & 4.24 & 6.22 & 6.16 & 6.50 & 9.79 \\ 
		\cline{2-8} 
		& GSAC & 3.39 & 3.90 & 6.07 & 4.59 & 5.14 & 9.13 \\ 
		\cline{2-8} 
		& Greedy & 2.82 & 2.83 & 3.35 & 4.37 & 4.61 & 5.66 \\ 
		\cline{2-8} 
		& Random & 1.65 & 2.09 & 5.76 & 1.45 & 1.71 & 7.56 \\ 
		\midrule
		\multirow{4}{*}{\begin{tabular}[c]{@{}c@{}}Cache-\\ hit Ratio\end{tabular}} & Optimal & 0.99 & 0.99 & 1.00 & 0.99 & 0.99 & 1.00 \\ 
		\cline{2-8} 
		& GSAC & 0.97 & 0.99 & 1.00 & 0.94 & 0.96 & 1.00 \\ 
		\cline{2-8} 
		& Greedy & 0.99 & 0.99 & 1.00 & 0.97 & 0.98 & 1.00 \\ 
		\cline{2-8} 
		& Random & 0.62 & 0.75 & 1.00 & 0.46 & 0.57 & 1.00 \\ 
		\bottomrule
	\end{tabular}
	}}
\end{table}

\subsection{Impact of the Prediction Accuracy ($\rho_{kn}$)}

\begin{figure}[htbp]
	\centering
	\begin{subfigure}{\figSizeOne\textwidth}
		\centering
		\includegraphics[width=\textwidth]{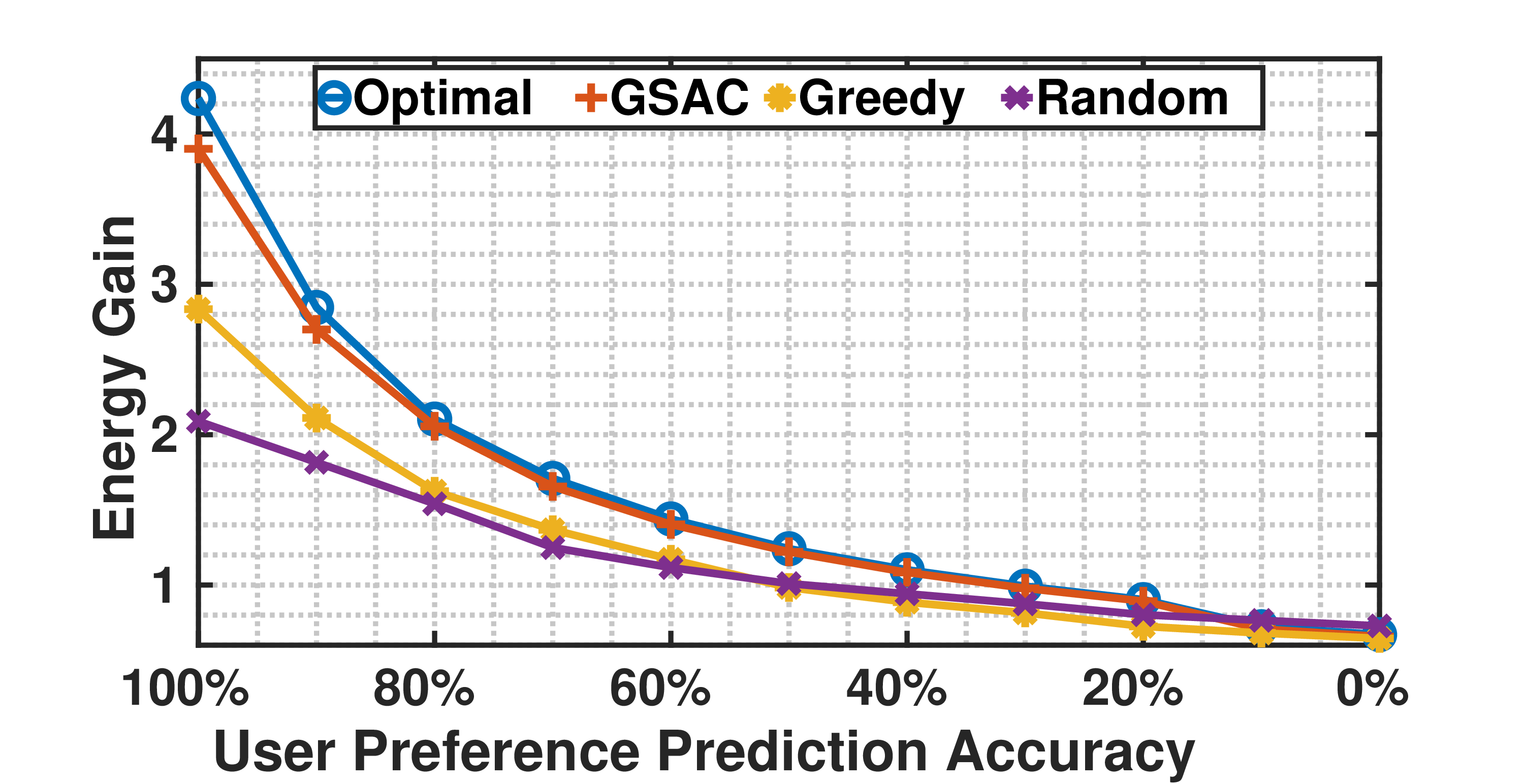}
		\caption{Energy Gain on P1}
		\label{fig:EG_P1_Acc}
	\end{subfigure}
	\begin{subfigure}{\figSizeOne\textwidth}
		\centering
		\includegraphics[width=\textwidth]{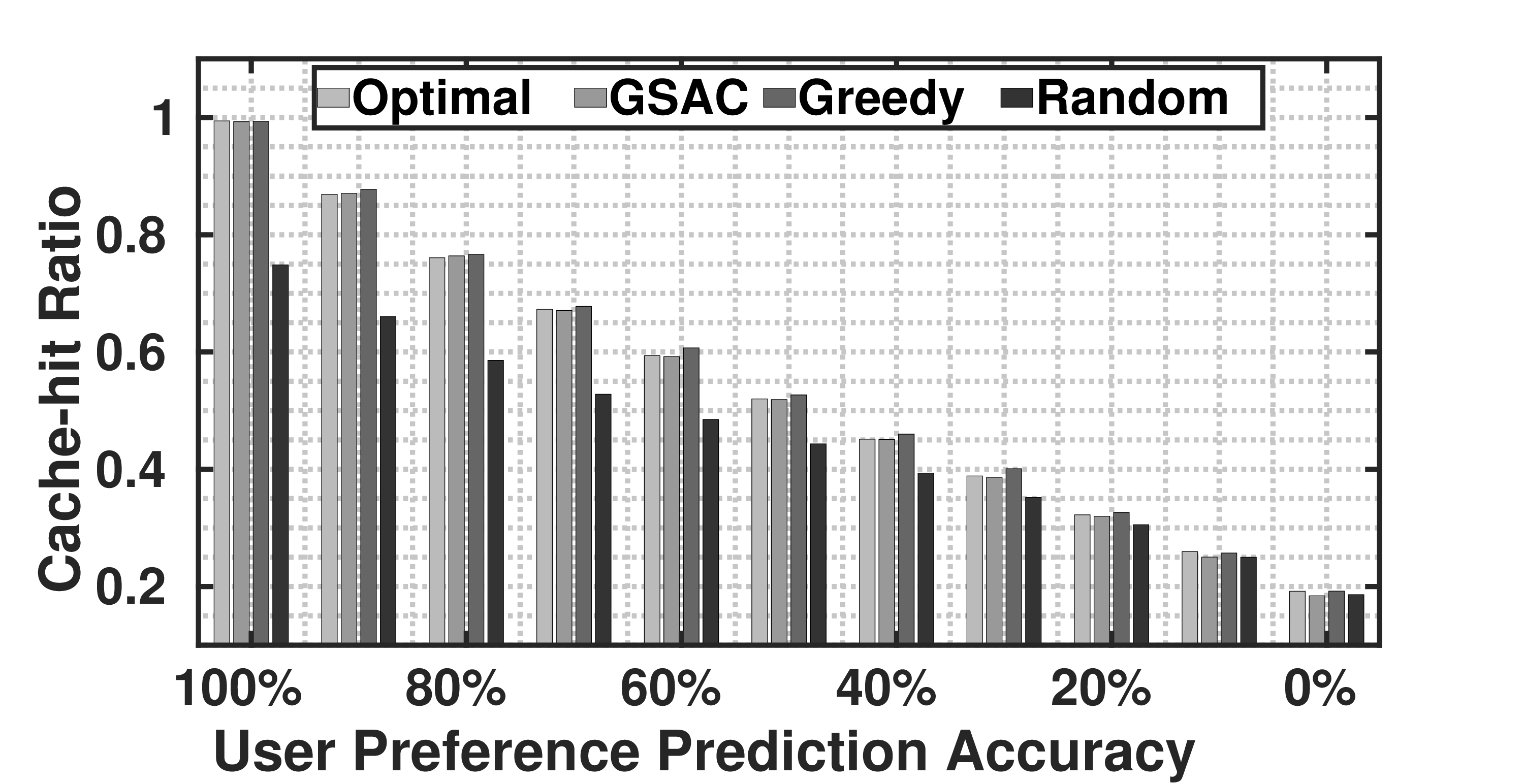}
		\caption{Cache-hit Ratio on P1}
		\label{fig:CR_P1_Acc}
	\end{subfigure}
	\begin{subfigure}{\figSizeOne\textwidth}
		\centering
		\includegraphics[width=\textwidth]{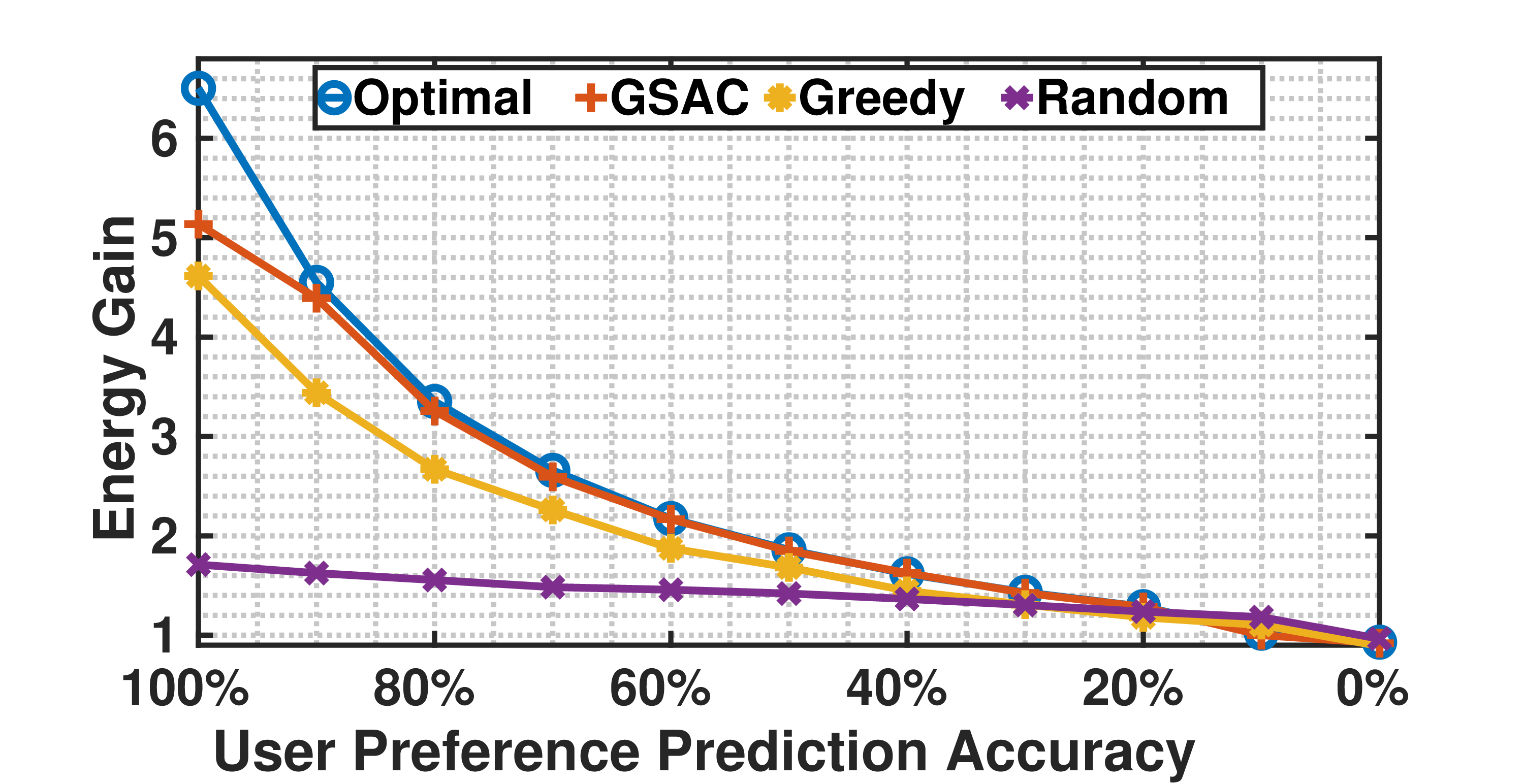}
		\caption{Energy Gain on P2}
		\label{fig:EG_P2_Acc}
	\end{subfigure}
	\begin{subfigure}{\figSizeOne\textwidth}
		\centering
		\includegraphics[width=\textwidth]{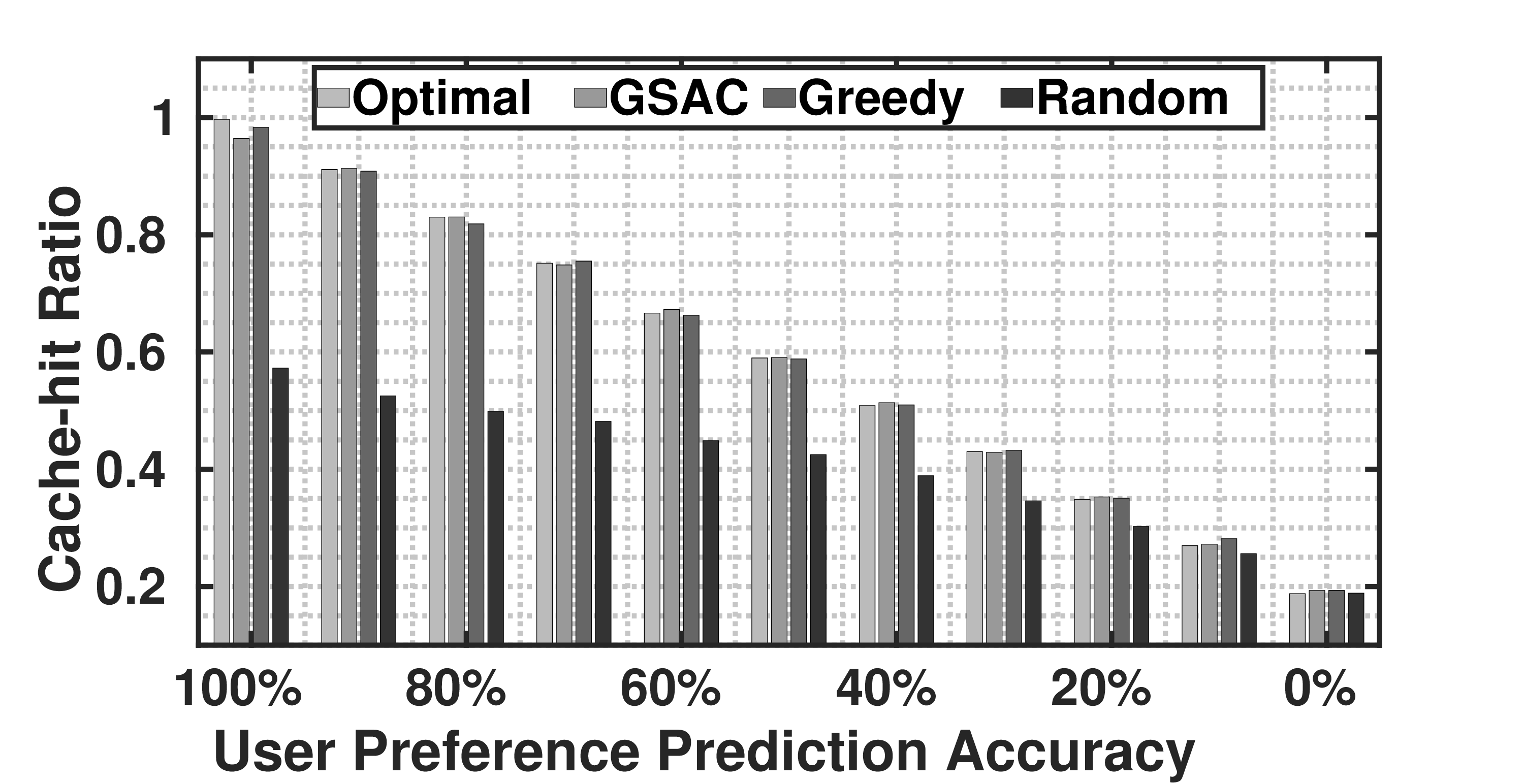}
		\caption{Cache-hit Ratio on P2}
		\label{fig:CR_P2_Acc}
	\end{subfigure}
	\caption{Algorithms Performance versus Preference Prediction Accuracy}
	\label{fig:per_accuracy}
\end{figure}

In the previous evaluation, we test the performance of different schemes under the salient assumption of an ideal condition that all users' preference is exactly predicted, i.e. $\rho_{kn}=1$. Hereafter, we relax this assumption and perform a wide set of experiments to understand the impact of prediction accuracy by varying it from $100\%$ down to $0\%$. As shown in Figure \ref{fig:per_accuracy}.(\subref{fig:EG_P1_Acc}) and \ref{fig:per_accuracy}.(\subref{fig:EG_P2_Acc}), regarding the Optimal, GSAC, and Greedy Caching, there is a steep drop on the energy gain from $100\%$ to $80\%$ then the value decline steadily. By the same token, the performance gap among the Optimal, GSAC, Greedy, and Random Caching becomes very small. It is worth noting that for $(P1)$, when the predicted accuracy is less than $30\%$ (For $(P2)$, the threshold is $10\%$), the energy gain is less than $1$. In other words, the proactive caching method cannot save energy but consume more than No-caching, if the user preference prediction is inaccurate, since we pay extra energy for caching contents. This result means that attention should be placed on managing effectively user content prediction in order to fully capitalize gains stemming from network and storage dimensioning.   In Figure \ref{fig:per_accuracy}.(\subref{fig:CR_P1_Acc}) and \ref{fig:per_accuracy}.(\subref{fig:CR_P2_Acc}), the cache-hit ratio reduces along with the accuracy declining. An interesting observation is that even when the prediction accuracy reduces to $0\%$ (for example in the case of very new popular content), the cache-hit ratio is still above $0$. The main reason is that inaccurate predicted requests can be redirected to other nodes where the required content is cached, and it is still counted for the cache-hit ratio calculation.  

\section{Conclusions}
\label{sec:conclusions}
To meet the significant growth of popular on-demand content whilst empowering a network energy efficiency operation, we jointly consider the problem of anchoring popular content at network nodes with caching capabilities and deciding delivery routing paths. The proposed integer linear programming model amalgamates energy consumption, the constraints of caching memory, link bandwidth, multipath routing, and different delivery modes, i.e. multicast and unicast. In addition, a greedy simulated annealing caching algorithm is presented and the effectiveness of the proposed schemes is evaluated by a wide set of numerical investigations. For instance, we test the performance of evaluated algorithms under different network resources, topologies, and user preference predictions. 
The large set of conducted experiments  suggests that the proposed algorithms have the potential to reduce the energy consumption up to $80\%$ compared with other caching schemes. Furthermore, expanding link bandwidth, enriching path diversity, and increasing the preference prediction accuracy have beneficial effects on energy saving. 
Furthering this work, we plan to consider ML/AI-driven algorithms amalgamated with  the proposed framework to augment the model with deep learning capabilities so that the optimal allocation patterns can be learnt autonomously. In that case a number of open ended questions arises such as the resource usage history to be used and the look-ahead horizon of the decision making actions. Those aspects open up new research avenues that may expand our contributions in future work.

\bibliographystyle{ieeetr}
\bibliography{references}

\end{document}